\documentclass[12pt]{article}
\pdfoutput=1
\usepackage{graphicx}
\usepackage{color}
\usepackage{cancel}
\usepackage{amssymb}
\usepackage{amsmath}
\usepackage{cite}
\usepackage{subfig}

\newcommand{\lsim}{\raisebox{-0.13cm}{~\shortstack{$<$ \\[-0.07cm] $\sim$}}~} 
 
\newcommand{\beq}{\begin{eqnarray}} 
\newcommand{\eeq}{\end{eqnarray}}

\begin{document}

\vspace*{5mm} 

\begin{center}

\mbox{\large\bf 
Charting Dark Matter down to the neutrino floor/fog}  

\vspace*{2mm}

\mbox{\large\bf in the 2HD+a scenario}

\vspace*{5mm}

{\sc Giorgio Arcadi}$^{1,2}$, {\sc Abdelhak~Djouadi}$^{3}$ and {\sc Stefano Profumo}$^{4}$ 

\vspace{5mm}

{\small 

$^1$ Dipartimento di Scienze Matematiche e Informatiche, Scienze Fisiche e Scienze della Terra, \\
Universita degli Studi di Messina, V. Ferdinando Stagno d'Alcontres 31, I-98166 Messina, Italy.\\[2mm]

$^2$ INFN Sezione di Catania, Via Santa Sofia 64, I-95123 Catania, Italy. \\[2mm]

\mbox{ \hspace*{-6mm} $^3$ Departamento de F\'isica Te\'orica y del Cosmos, Universidad de Granada,
18071 Granada, Spain.} \\[2mm] 

{\hspace*{-6mm} $^4$ Department of Physics and Santa Cruz Institute for Particle Physics (SCIPP),\\ University of California, Santa Cruz, Santa Cruz, CA 95064, USA.}
}

\end{center}

\vspace*{1cm}

\begin{abstract}
Next-generation direct detection experiments will probe dark matter (DM) scattering cross-sections deep into the neutrino fog, the regime where coherent neutrino scattering becomes an irreducible background. We investigate whether thermally produced weakly interacting massive particles (WIMPs) can naturally populate this regime while satisfying relic density and indirect detection constraints. Adopting as a case study the 2HD+a, a two-Higgs doublet model with {\it h,H,A,H}$^\pm$ scalar states supplemented by a light singlet pseudoscalar Higgs state ${\it a}$,   we have performed comprehensive parameter scans over the Type-I and Type-II Yukawa configurations. We have included the limit of strongly suppressed singlet--doublet {\it A-a} mixing sin$\theta \! \to \! 0$ and we show that annihilation into {\it ha} and ${\it Ha}$ final states sustains the correct relic
density while loop-induced direct detection cross-sections naturally land inside the neutrino fog; in the same limit the light pseudoscalar boson becomes long-lived, featuring displaced-vertex signatures when produced at colliders. Finally, in the case of zero mixing, we have considered a new possibility for DM phenomenology as the $a$ state becomes cosmologically stable and, consequently, an additional DM component. We map all the viable parameter space against current LZ and FERMI-LAT bounds and projected XLZD and CTA sensitivities. We find that the single component setup lies naturally below the neutrino floor for DM masses above 100 GeV while, on the contrary, most of the parameter space of the two component DM scenario is strongly disfavored already considering present limit. The parameter space of both single and two component DM scenario can be nevertheless broadened by considering specific relations among the model parameters to suppress the coupling between the 125 GeV bosons and two $a$ states. Our results represent in any case a motivation to fully exploit future tonne-scale detectors.
\end{abstract}

\newpage

\section{Introduction}
The nature of dark matter (DM) remains one of the central open questions in
particle physics and cosmology. Among the many proposed candidates, weakly
interacting massive particles (WIMPs) occupy a privileged position: their
thermal production via freeze-out in the early Universe naturally reproduces the
observed relic abundance \cite{Planck:2018vyg} for electroweak-scale masses and
couplings, a coincidence known as the WIMP miracle
\cite{Kolb:1990vq,Jungman:1995df,Bertone:2004pz}.
This theoretical appeal has driven a broad experimental program spanning
direct detection, indirect detection, and collider searches
\cite{Arcadi:2017kky,Arcadi:2024ukq,Arcadi:2019lka}.
 
Direct detection experiments search for the nuclear recoils produced by
DM scattering off target nuclei in underground detectors
\cite{Goodman:1984dc,Drukier:1986tm}.
Over the past decade this program has achieved enormous progress, with
tonne-scale liquid xenon detectors--PandaX-4T \cite{PandaX-4T:2021bab},
XENONnT \cite{XENON:2023cxc}, and most recently LUX-ZEPLIN (LZ) \cite{LZ:2022lsv}--
pushing spin-independent (SI) cross-section limits to few 
$10^{-48}\,\mathrm{cm^2}$ for DM masses near $40\,\mathrm{GeV}$.
The upcoming XLZD observatory \cite{XLZD:2024nsu} will extend sensitivity
by another order of magnitude.
 
This experimental progress, however, is approaching a fundamental barrier:
the neutrino fog \cite{Billard:2013qya,OHare:2021utq,Akerib:2022ort}.
The original literature referred to this as the ``neutrino floor,'' but
O'Hare~\cite{OHare:2021utq} showed that the concept is more precisely
captured as a fog: sensitivity does not vanish abruptly but degrades
continuously as experiments enter the cross-section region where
coherent elastic neutrino--nucleus scattering (CE$\nu$NS) from solar, atmospheric,
and diffuse supernova neutrinos produces nuclear recoil signals that
mimic those expected from DM.
Once an experiment accumulates sufficient exposure to observe $\mathcal{O}(1)$
neutrino scattering events, extracting a DM signal requires statistical
discrimination rather than simple counting, and the sensitivity gain per unit
exposure degrades significantly.
This defines the neutrino fog: the cross-section region, roughly
$\sigma_{\chi p}^{\rm SI} \lesssim 10^{-48}$--$10^{-47}\,\mathrm{cm^2}$
depending on DM mass, where next-generation experiments will begin to lose
the clean discovery power that has characterized earlier searches.
We use ``neutrino fog'' throughout this paper; where we write ``floor/fog''
in figure labels it refers to the same concept.
 
A physically important question is therefore whether viable WIMP candidates
can naturally predict scattering cross-sections inside the neutrino fog,
while simultaneously satisfying the relic density constraint and surviving
indirect detection limits from $\gamma$-ray, antiproton, and neutrino
observations \cite{Gaskins:2016cha}.
If they can, the neutrino fog is not the end of the WIMP story but rather
a motivation for new experimental and theoretical strategies.
 
One well-known mechanism that suppresses direct detection rates involves
pseudoscalar or axial-vector mediators between DM and 
quarks \cite{Dolan:2014ska,Arcadi:2017wqi}.
In the non-relativistic limit relevant for direct detection, pseudoscalar-mediated
interactions produce momentum-transfer-suppressed cross-sections, naturally
placing viable parameter space inside the neutrino fog without fine-tuning.
This suppression arises at tree level and is a generic feature of CP-odd mediators;
the thermal relic density and indirect detection signals are not similarly
suppressed, since DM annihilation in the early Universe and in DM halos proceeds
at higher velocities where $s$-wave contributions can dominate.
Pseudoscalar mediators thus generically predict a complementarity between
direct and indirect detection that deserves systematic study as experiments
approach the neutrino fog.

The most popular model embedding the aforementioned setup is the so called 2HD+a \cite{Bauer:2017ota,Ipek:2014gua,Arcadi:2022dmt,Arcadi:2022lpp},
which is currently the primary benchmark framework used by ATLAS and CMS for
DM searches at the LHC \cite{ATLAS:2021jbf,CMS:2021far}.
The model contains, besides two CP-even $h,H$ and two charged $H^\pm$ states from a two-Higgs doublet field,  a CP-odd singlet $a$ and a CP-odd doublet $A$ that mix
with angle $\theta$; in the limit $\sin\theta \to 0$ the direct detection
cross-section is entirely loop-induced, while annihilation into Higgs-associated
final states ($ha$, $Ha$) remains unsuppressed.
This limit is also of independent interest at the LHC, as the long lifetime
of the $a$ state produces displaced-vertex signatures
\cite{Haisch:2023rqs,Argyropoulos:2024yxo}.
 
Several previous studies have explored pseudoscalar-mediated DM at or near the
neutrino fog \cite{Arcadi:2017wqi,Abe:2018emu,Ertas:2019dew,Arcadi:2025sxc},
and the phenomenology of the 2HD+a has been investigated from many angles
\cite{Bauer:2017ota,Arcadi:2022dmt,Arcadi:2022lpp,Haisch:2023rqs,Argyropoulos:2024yxo,Arcadi:2020gge,Robens:2021lov,Argyropoulos:2022ezr}.
The present work makes three contributions that go beyond this literature.
\begin{itemize}
\item We perform the first unified parameter scan of the 2HD+a that extends to
strongly suppressed and vanishing $\sin\theta$, mapping the direct detection,
indirect detection, and LHC cross-sections simultaneously for all three mixing
regimes ($\sin\theta \geq 10^{-3}$, $\sin\theta < 10^{-3}$, $\sin\theta = 0$)
against current LZ and FERMI-LAT bounds and projected XLZD and CTA sensitivities.
\item In the $\sin\theta \to 0$ limit we establish a direct correlation between
sub-floor direct detection cross-sections and macroscopic $a$ lifetimes: the
same loop suppression that places DM inside the neutrino fog also generates
displaced hadronic signatures at the LHC with $c\tau_a$ values accessible to
existing ATLAS searches, making the two probes genuinely complementary rather
than redundant.
\item In the extreme $\sin\theta = 0$ limit we identify and characterize a
two-component DM (2CDM) scenario in which the pseudoscalar $a$ becomes cosmologically
stable. This implies a rather different phenomenology as the very different properties, under direct detection, of the pseudoscalar component with respect to the fermionic one. 
\end{itemize}
 
The paper is organized as follows.
Section~\ref{sec:2hdma} provides a general overview of the 2HD+a model and illustrates the general criteria of our study. The subsequent section is devoted to the main results, covering the standard
($\sin\theta \geq 10^{-3}$, Section~\ref{sec:2hdma_high}),
suppressed ($\sin\theta < 10^{-3}$, Section~\ref{sec:2hdma_low}),
and vanishing ($\sin\theta = 0$, Section~\ref{sec:2hdma_zero}) mixing
regimes in turn, and concluding with a discussion of LHC signatures
(Section~\ref{sec:2hdma_lhc}).\vspace*{-3mm}

%%%%%%%%%%%%%%%%%%%%%%%%%%%%%%%%%%%%%%%%%%%%%%%%%%

\section{The 2HD+a model: an overview}
\label{sec:2hdma}

The 2HD+a is a rather popular benchmark for the ongoing and future LHC searches of new physics and, at the same time,  features a very rich DM phenomenology with interesting connections, under a broader perspective, with cosmology. The model has been discussed at great length and details on its various aspects can be found e.g. in \cite{Bauer:2017ota,Ipek:2014gua,Arcadi:2022dmt,Arcadi:2022lpp,Robens:2021lov,Arcadi:2020gge,Argyropoulos:2022ezr,Argyropoulos:2024yxo}. Here, we simply summarize the main features of the model and briefly present the relevant theoretical and experimental constraints to which it is subject. 

The model is primarily characterized by a scalar potential
(that we will consider to be CP-conserving and with a soft $Z_2$ breaking version as it is customary) containing two SU(2) doublets of scalar fields $\Phi_{1,2}$ (2HDM) that develop vacuum expectations values $v_{1,2}$ and a CP-odd SU(2) singlet field. After electroweak  symmetry breaking, the physical spectrum that emerges is made of two CP-even bosons, dubbed $h,H$, with $h$ being the 125 GeV Higgs boson, a charged Higgs pair $H^{\pm}$, and two CP odd bosons, dubbed $a$ and $A$. 

The latter states are mixtures of the original singlet and doublet components weighted by a mixing angle $\theta$,   defined as
\begin{equation}
\left(
\begin{array}{c} A_0 \\ a_0 \end{array} \right)= \left( \begin{array}{cc} \cos\theta & \sin\theta \\ -\sin\theta & \cos\theta
\end{array} \right)
\left( \begin{array}{c} A \\ a \end{array}  \right)  
\ {\rm with} \ 
    \tan 2\theta = \frac{2 \kappa v_h}{M_A^2 - M_a^2}.
\end{equation}
with $v_h= (v_1^2+v_2^2)^{1/2} \simeq246\,\mbox{GeV}$ and $\kappa$ measuring the coupling between the doublet and singlet fields in the scalar potential  $V_{\rm 2HD+a}= V_{\rm 2HDM} + V_{\rm a_0}$ with the singlet component (the scalar potential of the 2HDM can be found in e.g. Ref.~\cite{Branco:2011iw}) given by
\begin{align}
      & V_{\rm a_0}= \frac{1}{2} m_{a_0}^2 a_0^2+ \frac{\lambda_a}{4} a_0^4+ \left(i \kappa a_0 \Phi^{\dagger}_1\Phi_2+\text{h.c.}\right)
    + \left[\lambda_{1P} a_0^2 \Phi_1^{\dagger}\Phi_1 \!+\! \lambda_{2P} a_0^2 \Phi_2^{\dagger}\Phi_2\right].\label{eq:V2HDMa}
\end{align}
Notice that, in the given parametrization, the mixing angle $\theta$ is defined in the range $\left[-\frac{\pi}{4},\frac{\pi}{4}\right]$ and the mass hierarchy $M_a < M_A$ automatically imposed.

The relevant phenomenology of the model is encoded in the following Lagrangian:
\begin{equation}
    \mathcal{L}_{\rm DM}+\mathcal{L}_{\rm Yuk}+\mathcal{L}_{\rm tri} \label{Lag-all}
\end{equation}
where $\mathcal{L}_{\rm DM}$ represents the interaction Lagrangian of the DM particle explicitly given by
\begin{equation}
    \mathcal{L}_{\rm DM}=i y_\chi \bar \chi \gamma_5 \chi (A \sin\theta+a \cos\theta) 
\end{equation}
with $\chi$ being a fermionic SU(2)-singlet DM candidate 
taken here to be of Dirac type (there is substantially no phenomenological difference if it were of the Majorana type) coupled to  two $A,a$ CP-odd neutral Higgs bosons. Next, we  have the Yukawa Lagrangian containing the coupling between the neutral Higgs bosons (we ignore, for simplicity, the ones of the charged Higgs states which can be found in Ref.~\cite{Arcadi:2022lpp}) and the standard fermions:
\begin{equation}
\mathcal{L}_{\rm Yuk}=\sum_f \frac{m_f}{v_h}\bigg[ g_{hff} h \bar f f+g_{Hff}
H\bar f f- i g_{Aff} A \bar f \gamma_5 f-i g_{aff} a \bar f \gamma_5 f  \bigg] \, , 
\end{equation}
The couplings are parameterized by applying scaling factors $g_{\phi ff}$ to the Standard Model (SM) Higgs Yukawa coupling $m_f/v_h$. These scaling factors are functions of the mixing angles $\beta$ and $\alpha$. The former derives from the ratio of vevs of the two-Higgs doublet fields, $\tan\beta =v_2/v_1$.
%with perturbativity requirements constraining it to lie (in general) in the range $\frac13 \lsim \tan\beta \lsim 50$. 
The angle $\alpha$ allows to diagonalize the 2HDM CP-even Higgs sector and which, in the so called alignment limit where the coupling of the lighter  $h$ state reaches it SM-like values, i.e. $g_{hff}\!=\!1$, reduces to $\alpha =\beta -\frac12 \pi$. In the pseudoscalar case, the couplings will depend also on the mixing angle $\theta$ since we have $g_{Aff}\!=\!\cos\theta g_{A^0ff}$ and $g_{aff}\!=\!\sin\theta \, g_{A^0ff}$. However, only four specific functional forms of the scaling parameters, prevent the appearance of flavor changing neutral current at tree-level \cite{Branco:2011iw}. These are dubbed, analogously to the conventional 2HDM, Type-I, Type-II, Type-X and Type-Y. One can then eliminate the dependence on the angle $\alpha$ by imposing the alignment limit, ensuring SM-like coupling for the $h$ state (i.e. $g_{hff}=1$). The Yukawa couplings of the $H$ state are the same as the $A^0$  ones in this limit. In the four possible configurations, along the lines of reasoning just illustrated above, the $A^0$ couplings are summarized in Table \ref{table:2hdm_cplgs}.

\begin{table}[ht!]
\vspace*{-.3cm}
\renewcommand{\arraystretch}{1.5}
\begin{center}
\begin{tabular}{|c|c|c|c|c|}
\hline
~~~~~~ &  ~~Type-I~~ & ~~Type-II~~ & ~~Type-X~~ &  ~~Type-Y~~ \\ \hline \hline 
$g_{A^0tt}$ & ${\cot\beta}$ & ${\cot\beta}$ & ${\cot\beta}$ & ${\cot\beta}$ \\ \hline
$g_{A^0bb}$ & $-{\cot\beta}$ & ${\tan\beta}$ & $-{\cot\beta}$ & ${\tan\beta}$ \\ \hline
$g_{A^0\tau\tau}$ & $-{\cot\beta}$ & ${\tan\beta}$ & ${\tan\beta}$ & $-{\cot\beta}$
\\ \hline
\end{tabular}
\vspace*{-1mm}
\caption{The Yukawa couplings of the pseudoscalar $A^0$ boson in the four Yukawa configurations; the of the CP-even $H$ are the same in the alignment limit $\alpha\!=\!\beta\!-\! \pi/2$.}
\label{table:2hdm_cplgs}
\end{center}
\vspace*{-.6cm}
\end{table}

The last relevant term in the Lagrangian of eq.~(\ref{Lag-all}) refers to the trilinear couplings of two CP-odd and one CP-even Higgs boson, namely:
\begin{equation}
\label{eq:lagrangian_trilinear}
    \mathcal{L}_{\rm tri}=\sum_{\phi=h,H}\lambda_{\phi aa} \phi aa+\lambda_{\phi Aa} \phi Aa+ \lambda_{\phi AA} \phi AA \, . 
\end{equation}
Analytical expressions of such couplings have been provided, mostly in the alignment limit, in e.g. \cite{Bauer:2017ota,Arcadi:2019lka,Robens:2021lov,Arcadi:2024ukq}. Being relevant for the discussion below, we provide here the expression of the $haa$ coupling in the alignment limit: 
%Using, for compactness, the abbreviations $s_X, c_X=\sin(X),\cos(X)$,  $t_\beta=\tan\beta$:
\begin{equation}
\label{eq:tril}
  \lambda_{haa}\!=\!\frac{1}{v_h}\left[\left(M_h^2\!+\!2 M_H^2\!-\!2 M_a^2\!-\!2 \lambda_3 v^2\right)\sin^2 \theta\!-\!2 \left(\lambda_{P1} \cos^2 \beta\!+\!\lambda_{P2}\sin^2 \beta\right)v^2 \cos^2 \theta \right] \, .
\end{equation}
The model features the following free parameters: the Higgs masses $M_H,M_{H^{\pm}},M_A,M_a$ ($M_h$ being fixed to 125 GeV), a 2HDM quartic coupling $\lambda_3$,  
and three quartic couplings involving $a$, $\lambda_{1P},\lambda_{2P},\lambda_a$, the mixing angles in the Higgs sector, $\alpha,\beta,\theta$ and, when the dark sector is included, the DM parameters $m_\chi,y_\chi$. We present an analysis of the 2HD+a based on wide parameter scans over the following parameter ranges:
%%%%%%%%%%%%%%%%%%%%%%%%%%%%%%%%%%%%%
\begin{align}
\label{eq:2HDMascan}
    & M_H,\, M_{H^{\pm}},\, M_A \in [M_h,1000]\,\mathrm{GeV},\quad M_a \in [10,600]\,\mathrm{GeV}, \notag\\
    & |\sin\theta| \leq 1/ \sqrt{2},\quad \tan\beta \in [1,10],\,\,\,\left \vert \cos(\beta-\alpha) \right \vert \leq 0.2 \, ,   \notag\\
    & \lambda_3,\, \lambda_{1P},\, \lambda_{2P} \in [-4\pi,4\pi], \quad m_\chi \in [1,1000]\,\mathrm{GeV},\quad y_\chi \in [10^{-3},10] \, .
\end{align}
%%%%%%%%%%%%%%%%%%%%%%%%%%%%%%%%%%%%%
As mentioned earlier, the most distinctive feature relative to existing literature is the range of values of $\sin\theta$, which varies from the maximal mixing, i.e. $\sin\theta\!= \!1/\sqrt{2}$ to the limit $\sin\theta \!\rightarrow \!0$. We also note that $\tan\beta$ can be actually varied, while remaining consistent with perturbative unitarity, between $1/3$ and $50$. We have chosen a more restricted range of variation to automatically account for some collider constraints (see below) and allow for a more straightforward comparison between different configurations of the Yukawa couplings. 

The scan has been performed for the Type-I and II configurations of the Yukawa couplings. Indeed, as pointed out e.g. in Ref.~\cite{Arcadi:2022lpp}, the Type-X (Type-Y) model has a very similar phenomenology, under the DM perspective, with the Type-I (Type-II) models (since the difference is simply due to the different couplings to leptons which do not matter in most cases). For each scan, the following list of different constraints have been applied. 

\underline{Theoretical constraints} on the quartic couplings of the 2HD+a scalar potential which can be translated into constraints on the Higgs boson masses and, in particular, their relative splittings. These constraints can be found e.g. in Ref.~\cite{Arcadi:2022lpp} and mainly originate from the requirement of perturbative unitarity, which leads to bounds on the combinations of the self-couplings $\lambda_i (\lsim 4\pi)$, and a scalar potential that should be bounded from below, which forces some of the  couplings or their combinations to be positively defined. When combining these two requirements, one obtains limits on the mass splitting of the Higgs bosons, in particular of the doublet-like states $H,H^{\pm},A$. The other pseudoscalar $a$ can be instead lighter, possibly even lighter than the $h$ state, provided that $M_A$ does not exceed the value of around 1.4 TeV (this limit corresponds to the maximal mixing and can be weakened by decreasing $\sin 2 \theta$ and possibly disappears for $\sin\theta=0$). 

 \underline{Electroweak precision measurements}, we have considered possible deviation, with respect to the SM expectation, of the $S,T,U$ parameters, due to the extra Higgs bosons in the scalar sector. We have imposed that such deviations to be within the experimental limit via the same procedure illustrated in \cite{Arcadi:2022lpp}. The extra Higgs sector mostly affects that $T$ parameter (and similarly the so-called $\rho$ parameter  which measures the strength of the neutral to charged currents ratio at zero-momentum transfer). The SM value can be exactly recovered by imposing the $M_H=M_{H^{\pm}}=M_A$.  
 
\underline{Constraints from flavor physics} and, in particular, those coming from $b\rightarrow s \gamma$ radiative transitions. For the ranges of values of $\tan\beta$ considered in this study, the Type-II model is mostly affected with a lower bound of 800 GeV \cite{Misiak:2020vlo} on the mass of the charged Higgs boson. The same bound applies in the Type-I configuration for low $\tan\beta$ values.  In addition, the mode $B_s \to \mu^+\mu^-$ can potentially receive large contributions from the exchange of a light $a$ state if it has large couplings to $b$-quarks and muons as is the case in the Type-II model at high $\tan\beta$ values for instance. 

\underline{Constraints from Higgs searches at the LHC}. In particular, four types are relevant.\vspace*{-2mm} 

\begin{itemize}

\item Limits from the SM-like Higgs signal strength: for arbitrary values of $\alpha$ and $\beta$, the coupling of the $h$ boson deviate with respect to the values predicted in the SM. This possibility has been widely investigated by the ATLAS and CMS collaborations \cite{ATLAS-new,CMS-new} in the various Higgs production and decay processes, leading to very strong constraints. In particular, the Type-II setup is compatible only with the alignment limit $\beta-\alpha=\pi/2$ while for Type-I, one should have $\vert \cos(\beta-\alpha) \vert \leq 0.2$. This motivates the choice of the ranges, for $\cos(\beta-\alpha)$, in the scan \ref{eq:2HDMascan}.\vspace*{-2mm} 

\item Search for the heavy 2HDM $H,A$ and $H^\pm$ states: 
the most important channel is neutral Higgs production in gluon-fusion (via heavy quark loops) and their decay into the clean $\tau^+\tau^-$ final states $pp\to gg \to H/A \to \tau^+\tau^-$ which constrains large $\tan\beta$ values in the Type-II scenario or the decay into top quarks, $pp\to gg \to H/A \to t\bar t$ which constrains low $\tan\beta$ values in both Type-I and II configurations. $H/A$ masses close to 1 TeV and below are excluded in general.  Additional but less severe constraints come from associated $H^\pm$ production with top quarks. As already pointed out, such constraints have been mostly accounted for via the specific choice of the range of variation for $\tan\beta$.\vspace*{-2mm}  

\item Invisible decays of the SM-like $h$ boson: if $M_a\! \leq \! \frac12 M_h$ the decay $h \! \rightarrow \! aa$ is accessible and occurs with a rate $\Gamma \!=\! |\lambda_{haa}|^2/(32 \pi M_h) \! \times\! (1- 4 M_a^2/M_h^2)^{1/2}$. This possibility is severely tested at the LHC both via fits of the $h$ signal strengths \cite{ATLAS:2023tkt,CMS:2022qva} and via direct searches of the topology signatures associated to this decay process \cite{ATLAS:2025qyn,ATLAS:2024nnm,ATLAS:2021hbr,ATLAS:2020ahi,CMS:2025hjt,CMS:2024uru}. Since no signal has been detected so far, the possibility $M_a < \frac12 M_h$ is strongly disfavored, unless the coupling $\lambda_{haa}$ is very small, namely $\left \vert \lambda_{haa} \right \vert /M_h \leq 10^{-3}$. As can be seen from eq.~(\ref{eq:tril}), such condition might be realized by fine-tuning the $M_{H,A,H^{\pm},a}$ and $\lambda_{3,1P,2P}$ parameters. We have explicitly verified that, performing a general scan, as the one depicted in eq.~(\ref{eq:2HDMascan}), no viable point with $M_a \leq M_h/2$ was found. We could nevertheless extend the viable region to light $M_a$ by performing an additional scan in which the coupling $\lambda_{1P}$, in spite of being varied over a flat distribution of values, has been fixed, as function of $M_{H,a,A,H^{\pm}},\lambda_3,\lambda_{2P},\alpha,\beta,\theta$ to have $\left \vert \lambda_{haa} \right \vert/M_h \leq 10^{-3}$.\vspace*{-2mm} 

\item Single production of the $a$ state: with the strongest constrain being due to the processes in which it is produced singly in gluon or b-quark fusion and decays into muons,  $p  p \to gg, b\bar b \to a \to \mu^+ \mu^-$ \cite{Argyropoulos:2022ezr}. This is particularly the case if $a$ has a significant mixing  with the heavier $A$ state.
Outside the alignment limit and for a light $a$ boson,  processes like $gg\to h \to aZ$ or $q \bar q \to ha$ could be relevant.\vspace*{-2mm}  
\end{itemize}

Finally, we have the \underline{cosmological and astrophysical constraints} and mainly:\vspace*{-2mm}  
\begin{itemize} 
\item The requirement of a correct cosmological relic density, namely that in the conventional freeze-out mechanism that we assume, the experimentally favored value, $\Omega_{\chi}h^2 = 0.12 \pm 0.001$, is achieved in the DM pair annihilation processes. The most relevant final states are the SM fermionic ones that occur via $a/A$ boson exchange, $\chi \chi \to a^*, A^* \to \tau^+ \tau^-,  b\bar b, t \bar t$ but those involving Higgs bosons, such  as $\chi \chi \to a^*, A^* \to   h a , Zh$ or  $\chi \chi \to aa$ could also be relevant.\vspace*{-2mm} 

\item For $\sin\theta \ll 1$, the $a$ state might become very long lived and might lead to some astrophysical complications such as for the Big Bang Nucleosynthesis.  To avoid such problems,  we have imposed a lower bound on its lifetime, $\tau_a \geq 1\,\mbox{s}$.\vspace*{-2mm} 

\item Constraints from indirect detection (ID) of the DM. We have focused on the gamma-ray signal adopting as a reference limit, the most stringent constraint which, to our best knowledge, is given in Ref.~\cite{McDaniel:2023bju} in searches of signals in Dwarf Spheroidal Galaxies (DSph) by FERMI-LAT. We have also considered the expected sensitivity of the future Cherenkov Telescope Array (CTA), as determined in Ref.~\cite{CTA:2020qlo}\footnote{Note that the comparison between FERMI-LAT and CTA limits is not trivial as the former relies on DSph while the latter on searches of signals from the galactic center. The different sensitivities are hence also affected by astrophysical inputs as for example the J-factors. Along the same spirit as \cite{Arcadi:2024ukq} we wanted just to provide a general insight of current limits and the one reachable in a foreseeable future.}. Other signatures, as e.g. antiprotons, can provide potentially competitive constraints with the one reported here. We omitted them for simplicity and because they are strongly affected by uncertainties related to the propagation models and by systematics.\vspace*{-2mm} 

\item Finally, there are constraints from direct detection (DD) in DM elastic scattering on nuclei. For our study, we have adopted  the most recent limits of  the LZ collaboration \cite{LZ:2024zvo}, complemented by the future prospects from the XLZD experiment \cite{XLZD:2024nsu}. 

\end{itemize}

\section{Collider and DM phenomenology of the model}

To ease the understanding of the results we will distinguish three regime of the 2HD+a which depend on the value of the mixing angle between the $a$ and $A$ states, $\sin\theta$.\vspace*{-2mm} 

\subsection{\boldmath$\sin\theta \geq 10^{-3}$}
\label{sec:2hdma_high}

This is the regime considered in most of the existing 2HD+a literature.
As discussed in the introduction, the momentum-transfer suppression of
pseudoscalar-mediated interactions in the non-relativistic limit means that
tree-level DM--nucleon scattering is helicity-suppressed; the dominant
contribution to the SI cross-section arises instead at one loop via box and
triangle diagrams with virtual $a$, $A$, and CP-even bosons $h$, $H$ running
in the loops.
The full loop-induced SI cross-section on protons reads
\cite{Abe:2018emu,Ertas:2019dew}:
\begin{align}
\label{eq:2HDMa_full_loop}
&\sigma_{\chi p}^{\rm SI} = \frac{\mu_{\chi p}^2}{\pi}\frac{m_p^2}{v_h^2}
\Bigg|
  \sum_q f_q \sum_{\phi=h,H} \frac{g_{\phi qq} m_q}{v_h M_\phi^2} C_{q}^{\rm triangle}\nonumber\\
&+ \sum_{q=u,d,s} f_q C_{1,q}^{\rm box} + \sum_{q=u,d,s,c,b} \frac{3}{4}
  \bigl(q(2)+\bar{q}(2)\bigr)
  \bigl[C_{1,q}^{\rm box} + m_\chi C_{2,q}^{\rm box}\bigr]
+ \frac{2}{27} f_{TG} C_G^{\rm box}
\Bigg|^2,
\end{align}
where $f_q$ and $f_{TG}$ are the nucleon matrix elements, $q(2)$ and $\bar{q}(2)$
are twist-2 parton distribution moments, and $\mu_{\chi p}$ is the
DM–proton reduced mass.
The Wilson coefficients $C_q^{\rm triangle}$, $C_{1,q}^{\rm box}$,
$C_{2,q}^{\rm box}$, and $C_G^{\rm box}$ encode the triangle, box, and gluon-fusion topologies respectively. Their expressions have been recently collected in \cite{Arcadi:2025sxc}.

The DM relic density is computed numerically with
\textsc{micrOMEGAs} \cite{Belanger:2006is,Alguero:2023zol}.
Analytic insight follows from the three dominant annihilation channels.
At leading order in the velocity expansion \cite{Arcadi:2019lka,Abe:2018emu,Arcadi:2022lpp,Arcadi:2024ukq}:
\begin{align}
\label{eq:2HDMa_sigmav}
\langle\sigma v\rangle_{\bar{f}f} &=
  \frac{1}{2\pi}\sum_f n_f^c \sqrt{1-\frac{m_f^2}{m_\chi^2}}\,
  y_\chi^2\sin^2\!\theta\cos^2\!\theta\, m_\chi^2
  \left|\frac{1}{4m_\chi^2-M_a^2}-\frac{1}{4m_\chi^2-M_A^2}\right|^2,
  \nonumber\\
\langle\sigma v\rangle_{ha} &=
  \frac{y_\chi^2}{16\pi}
  \sqrt{1-\frac{(M_h+M_a)^2}{4m_\chi^2}}
  \sqrt{1-\frac{(M_h-M_a)^2}{4m_\chi^2}}
  \left|\frac{\lambda_{haa}\cos\theta}{4m_\chi^2-M_a^2}
       +\frac{\lambda_{hAa}\sin\theta}{4m_\chi^2-M_A^2}\right|^2,
  \nonumber\\
\langle\sigma v\rangle_{aa} &=
  \frac{v_{\rm rel}^2}{12\pi}
  \left(1-\frac{M_a^2}{m_\chi^2}\right)^{5/2}
  y_\chi^4\,\frac{m_\chi^6}{(M_a^2-2m_\chi^2)^4}.
\end{align}
The $\bar{f}f$ channel is $s$-wave dominated and proportional to
$\sin^2\!\theta\cos^2\!\theta$; it controls the relic density for
$m_\chi \lesssim \frac12 (M_h+M_a)$ and gives the thermal cross-section
value probed by ID.
The $ha$ channel is also $s$-wave and becomes the dominant annihilation
mode at higher masses once it is kinematically accessible, due to the
potentially large trilinear coupling $\lambda_{haa}$.
The $aa$ channel is $p$-wave suppressed and contributes significantly
only for $y_\chi \gtrsim 1$.
The hierarchy between the $\bar{f}f$ and $ha$ channels inverts sharply
at the kinematic threshold $m_\chi = (M_h+M_a)/2$: below threshold the
$\bar{f}f$ rate carries a $\sin^2\!\theta\cos^2\!\theta \leq 1/4$
prefactor and the $ha$ channel is simply closed, while above threshold
$\langle\sigma v\rangle_{ha}$ is governed by $\lambda_{haa}$, which can
reach $\mathcal{O}(1)$--$\mathcal{O}(10)$ in units of $v_h$ for natural
quartic couplings, easily outpacing the $\sin^2\!\theta$-suppressed
$\bar{f}f$ rate across much of parameter space.
This inversion is the origin of the $m_\chi \gtrsim 100\,\text{GeV}$
preference visible in the scan results discussed below.

%%%%%%%%%%%%%%%%%%%%%%%%%%%%%%%%%%

\begin{figure}
    \centering
    \includegraphics[width=0.49\linewidth]{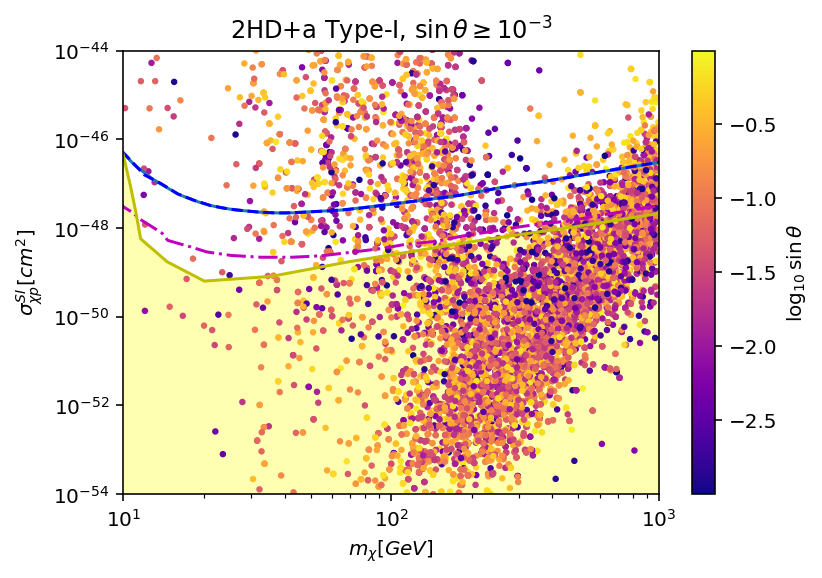}
    \includegraphics[width=0.49\linewidth]{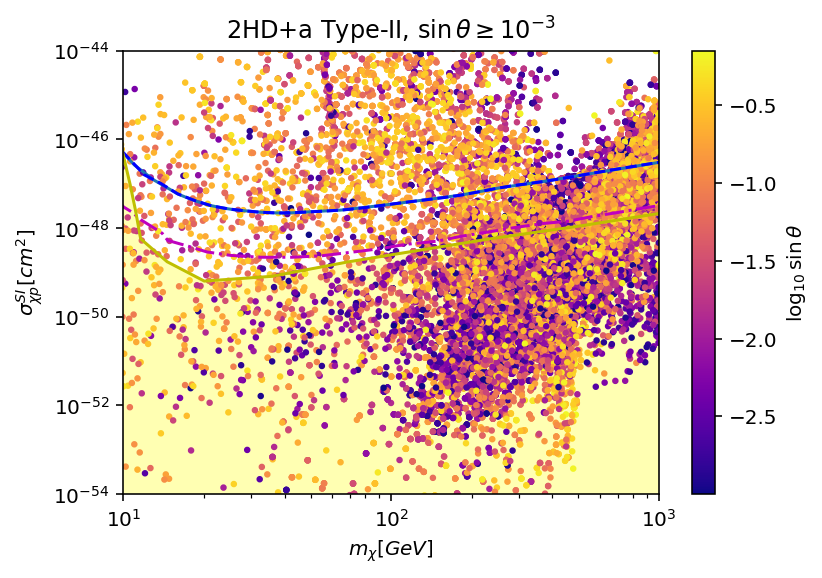}\\
    \includegraphics[width=0.49\linewidth]{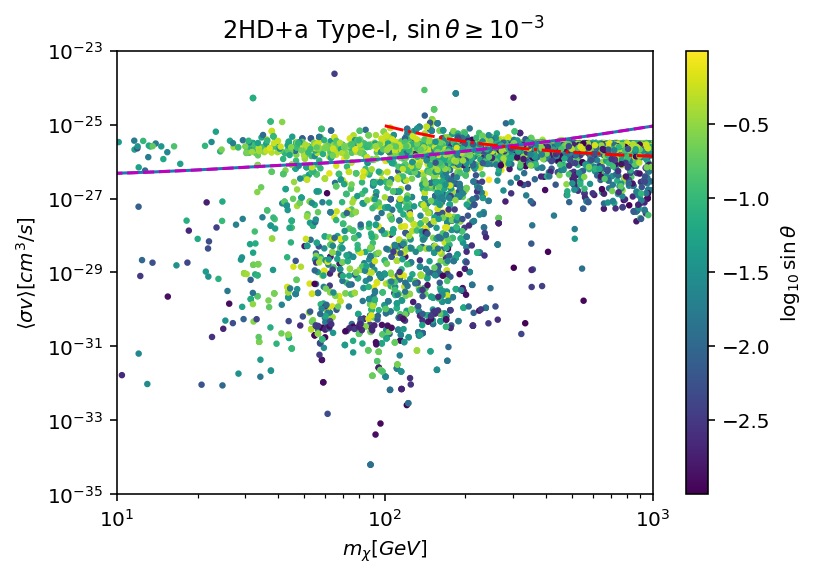}
    \includegraphics[width=0.49\linewidth]{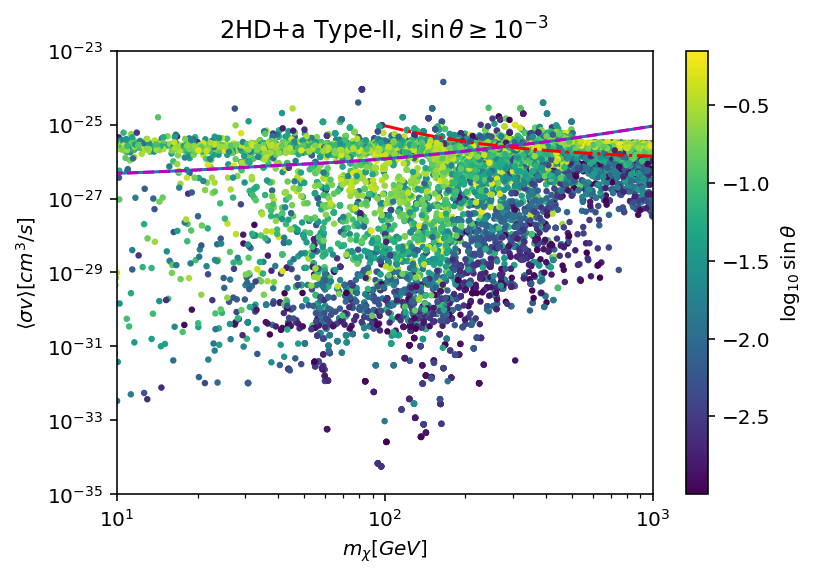}
    \vspace*{-2mm}
    \caption{Model points from parameter scans (see main text for details) of the 2HD+a reported in the $(m_\chi,\sigma_{\chi p}^{\rm SI})$ plane (upper row) and $(m_\chi,\langle \sigma v \rangle)$ plane (bottom row). The first column corresponds to the Type-I configuration of the coupling of the Higgs bosons to the SM fermions while the second column refers to the same for the Type-II. In the plots referring to DD, the solid (dashed) lines represent the current (future) sensitivity of direct detection experiments represented by LZ (XLZD) while the yellow region corresponds to the neutrino fog. In the case of ID, the current constraint from searches of DSph by FERMI-LAT and the future prospects by CTA are shown. In each panel, the color pattern of the model points correspond to the value of $\sin\theta$.}
    \label{fig:plotth1}
\vspace*{-3mm}
\end{figure}

The limits and prospects for direct and indirect detection are illustrated in Fig.~\ref{fig:plotth1},
which shows parameter assignments satisfying all listed constraints in the
$(m_\chi,\sigma_{\chi p}^{\rm SI})$ and $(m_\chi,\langle\sigma v\rangle)$ planes.
Following the discussion of the previous section, current LZ bounds and projected XLZD sensitivity delimit the direct detection reach; the neutrino fog occupies the yellow region.
For indirect detection we show the current FERMI-LAT and the projected CTA sensitivity.
The density of points inside the yellow region confirms that a large fraction of
the thermally viable 2HD+a parameter space falls naturally inside the neutrino fog,
motivating deeper experimental exploration of this regime.

The comparison of the two rows of figure~\ref{fig:plotth1} shows the DD/ID
complementarity operating in practice: combined constraints favor
$m_\chi \gtrsim 100\,\text{GeV}$, particularly in the Type-I scenario.
The SI cross-section spans several orders of magnitude for fixed $m_\chi$,
with the bulk of viable points lying near or below the current LZ exclusion
for large $\sin\theta$, and sinking towards the neutrino fog for small
$\sin\theta$ where the loop suppression is more complete.

\begin{figure}
    \centering
    \includegraphics[width=0.49\linewidth]{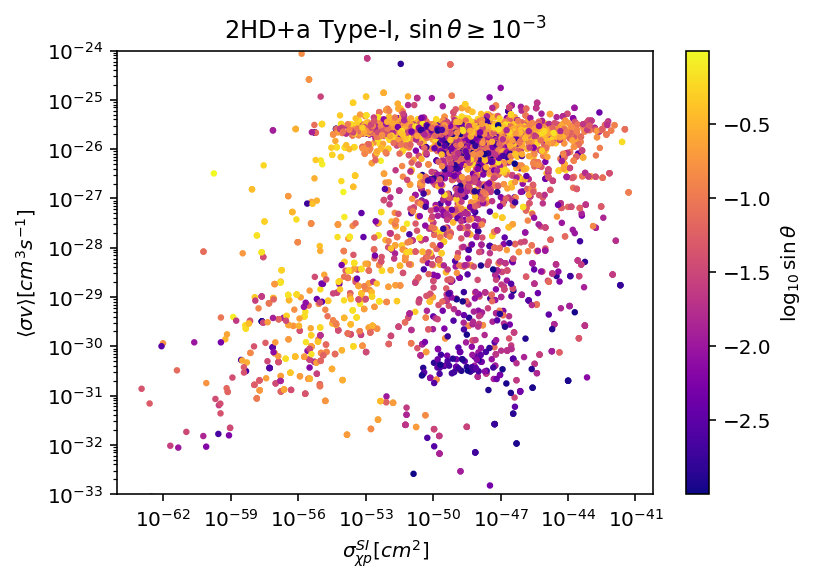}
    \includegraphics[width=0.49\linewidth]{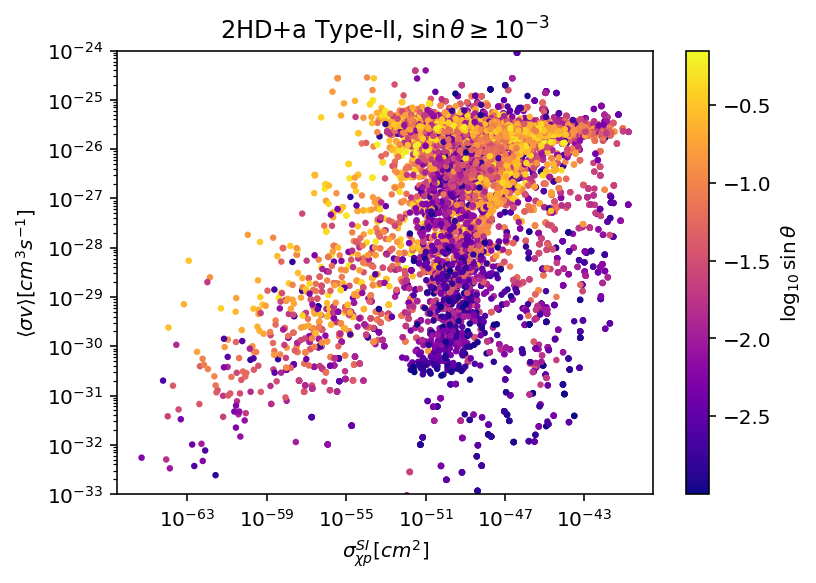}
    \caption{Same model points displayed in the upper row of Fig. \ref{fig:plotth1}. The color pattern this time correspond to the value of the DM annihilation cross-section normalized to $10^{-26}\,\mbox{cm}^3 {\mbox{s}}^{-1}$.}
    \label{fig:plotthF}
\end{figure}

Figure~\ref{fig:plotthF} re-displays the same model points in the
$(\sigma_{\chi p}^{\rm SI}, \langle\sigma v\rangle)$ plane, with color now
encoding $\langle\sigma v\rangle / (10^{-26}\,\text{cm}^3\,\text{s}^{-1})$.
This projection makes the direct/indirect detection complementarity explicit:
parameter points that evade current DD limits via loop suppression tend to
have an $s$-wave annihilation cross-section near or above the thermal value,
placing them within the reach of next-generation gamma-ray experiments.

\subsection{\boldmath$\sin\theta < 10^{-3}$, $\sin\theta \neq 0$}
\label{sec:2hdma_low}

We now ask whether viable DM parameter space extends to strongly suppressed
values of the singlet--doublet mixing.
This regime is of interest both for DM phenomenology and for LHC searches:
for $\sin\theta \ll 1$ the pseudoscalar $a$ acquires a macroscopically long
lifetime and can produce displaced-vertex signatures at colliders \cite{Haisch:2023rqs}.

The relic density constraint can still be satisfied in this regime.
Inspecting eq.~(\ref{eq:2HDMa_sigmav}), the annihilation cross-section into
$\bar{f}f$ is proportional to $\sin^2\!\theta\cos^2\!\theta$ and becomes
negligible for $\sin\theta \!\to \!0$, except near the $s$-channel poles
$m_\chi \!\simeq \!\frac12 M_{a,A}$.
However, the $ha$ and $Ha$ channels remain unsuppressed because the relevant
trilinear couplings are non-zero at $\sin\theta \!=\! 0$:
\begin{align}
\label{eq:lhaa_limit}
\lambda_{haa}\big|_{\sin\theta=0} &= -\frac{2v_h}{M_h}
  \left(\lambda_{1P}\cos^2\beta + \lambda_{2P}\sin^2\beta\right), \\
\label{eq:lHaa_limit}
\lambda_{Haa}\big|_{\sin\theta=0} &= \frac{v_h}{M_H}\sin 2\beta
  \left(\lambda_{1P} - \lambda_{2P}\right).
\end{align}
Similarly, the $aa$ annihilation cross-section is proportional to
$\cos^2\!\theta \to 1$ and remains fully operative.
Near the poles $m_\chi \simeq M_{a,A}/2$, one must also account for the
effective loop-induced coupling of DM pairs to the CP-even Higgs bosons that
is already responsible for the SI cross-section; this contribution was first
computed for a single pseudoscalar mediator in \cite{Belyaev:2022qnf} and
we have adapted it to the 2HD+a model.

\begin{figure}
    \centering
    \includegraphics[width=0.49\linewidth]{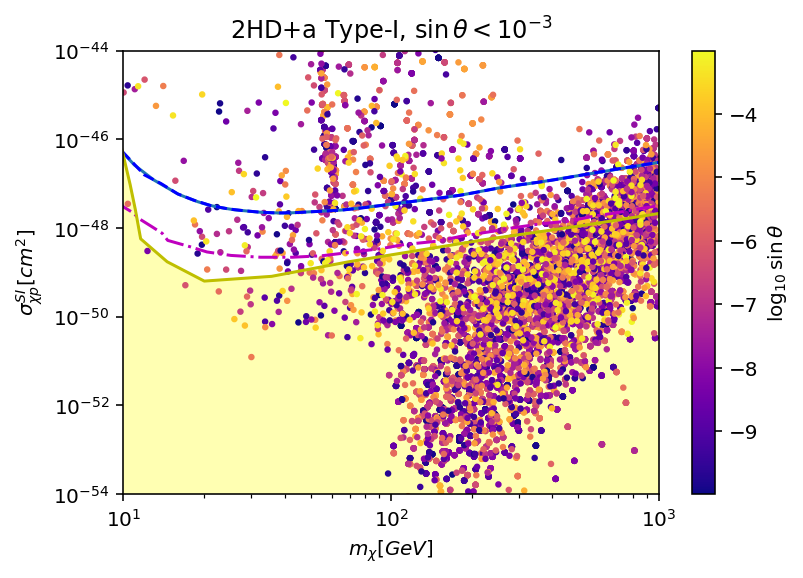}
    \includegraphics[width=0.49\linewidth]{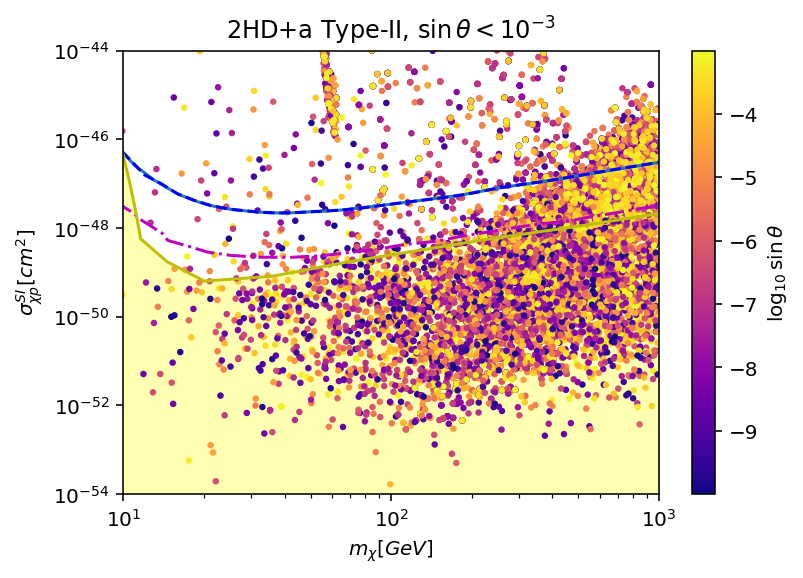}\\
    \includegraphics[width=0.49\linewidth]{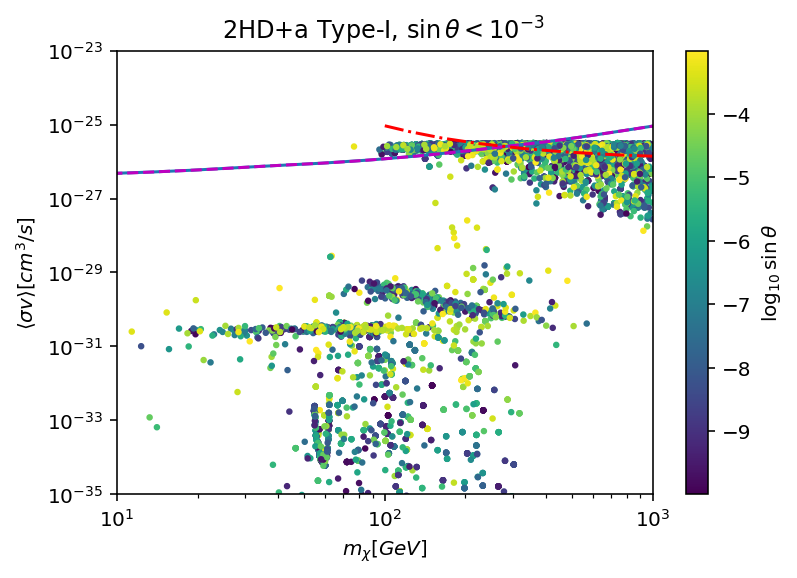}
    \includegraphics[width=0.49\linewidth]{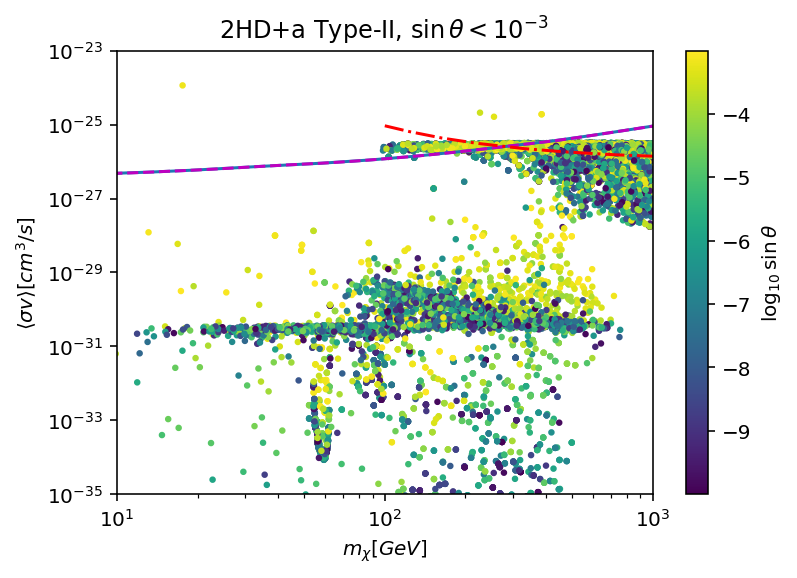}
    \caption{Same as Fig. \ref{fig:plotth1} but for the model points with $\sin\theta < 10^{-3}$.}
    \label{fig:plotth2}
\end{figure}

The corresponding DD and ID prospects are shown in Fig.~\ref{fig:plotth2},
in the same format as Fig.~\ref{fig:plotth1}.
Comparing the two figures, the low-$\sin\theta$ scan shows a higher density
of viable points at $m_\chi \gtrsim 100\,\text{GeV}$, confirming that the
$ha$ (and, for $\lambda_{1P} \neq \lambda_{2P}$, $Ha$) channel is the primary
freeze-out mechanism.
A secondary cluster appears near $m_\chi \simeq 60\,\text{GeV}$, corresponding
to the Higgs-pole region $m_\chi \simeq M_h/2$, where annihilation proceeds
via the loop-induced $\bar\chi\chi h$ vertex.
Analogously to the pure Higgs portal \cite{Arcadi:2019lka,Ballesteros:2020adh,Arcadi:2024ukq},
this pole region is strongly disfavored by DD constraints.

Especially in the Type-II scenario, the viable parameter region at low DM masses
can be extended by suppressing the $\lambda_{haa}$ coupling.
From eq.~(\ref{eq:lhaa_limit}), in the $\sin\theta \to 0$ limit this requires
$\lambda_{1P} \simeq -\lambda_{2P}\tan^2\!\beta$, a relation that is not
enforced by any symmetry of the model and therefore constitutes a tuning of
the quartic couplings at the level of $\mathcal{O}(\tan^2\!\beta)$.
To map this region we perform a dedicated subsidiary scan in which $\lambda_{1P}$
is fixed, as a function of $M_{H,a,A,H^\pm}$, $\lambda_{3}$, $\lambda_{2P}$,
$\alpha$, $\beta$, and $\theta$, to satisfy $|\lambda_{haa}|/M_h \leq 10^{-3}$,
analogously to the procedure already applied for the light-$a$ constraint
in section~\ref{sec:2hdma}.
In this tuned sub-space the $ha$ freeze-out channel is kinematically or
dynamically suppressed, and the correct relic density is achieved primarily
via the $p$-wave process $\bar\chi\chi \to aa$, which requires $y_\chi \gtrsim 1$
to compensate the velocity suppression.
The resulting scan points populate the low-mass region of
Fig.~\ref{fig:plotth2} that would otherwise be absent; they form a
recognizable sub-population at large $y_\chi$ and small $M_a$, and are
predominantly inside the neutrino fog given the further loop suppression
of the SI cross-section at vanishing $\lambda_{haa}$.

\begin{figure}
    \centering
    \includegraphics[width=0.49\linewidth]{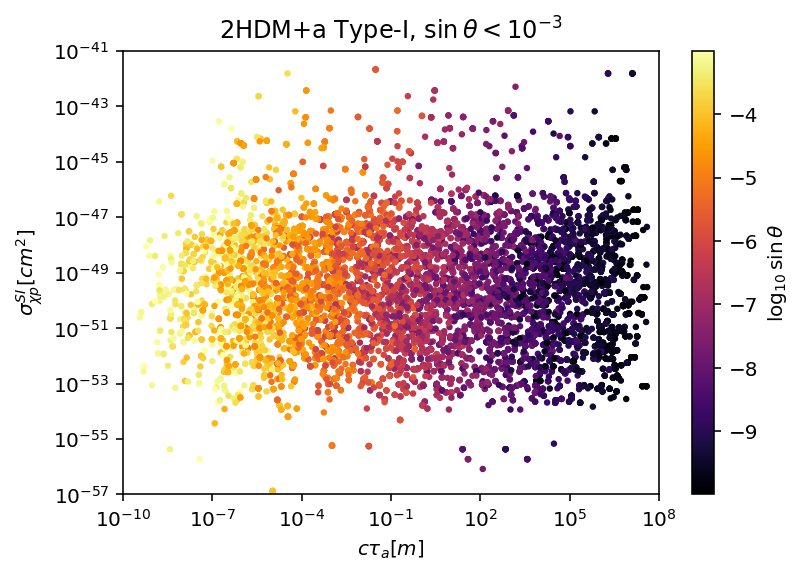}
    \includegraphics[width=0.49\linewidth]{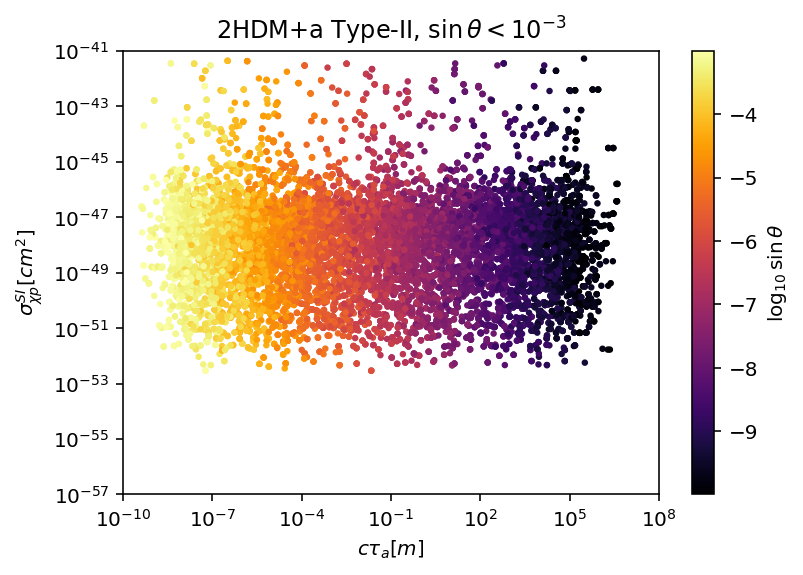}
    \caption{Model points, in the $\sin\theta < 10^{-3}$ regime, featuring the correct DM relic density and complying at the same time with current constraints from direct and indirect detection. The two panels correspond to, respectively, the Type-I and Type-II scenarios. In each panel, the color pattern of the model points tracks the decay length, $c \tau_a$ of the $a$ state.}
    \label{fig:2HDMa_tau}
\end{figure}

The connection to LHC phenomenology is illustrated in Fig.~\ref{fig:2HDMa_tau},
which shows the surviving model points in the $(\sigma_{\chi p}^{\rm SI},\, c\tau_a)$
plane, with color encoding $\sin\theta$.
The decay length $c\tau_a$ spans from sub-millimeter to cosmological scales over
the parameter space, and much of the region compatible with current DD constraints
or inside the neutrino fog corresponds to $c\tau_a$ values accessible to
existing and planned LHC searches for displaced hadronic jets
\cite{ATLAS:2018tup,ATLAS:2019qrr,ATLAS:2019jcm,ATLAS:2022gbw,ATLAS:2022zhj}.
This demonstrates a genuine complementarity: points that are difficult to reach
via conventional DM experiments may be accessible at the LHC through long-lived
particle signatures. 

\begin{figure}
    \centering
    \includegraphics[width=0.49\linewidth]{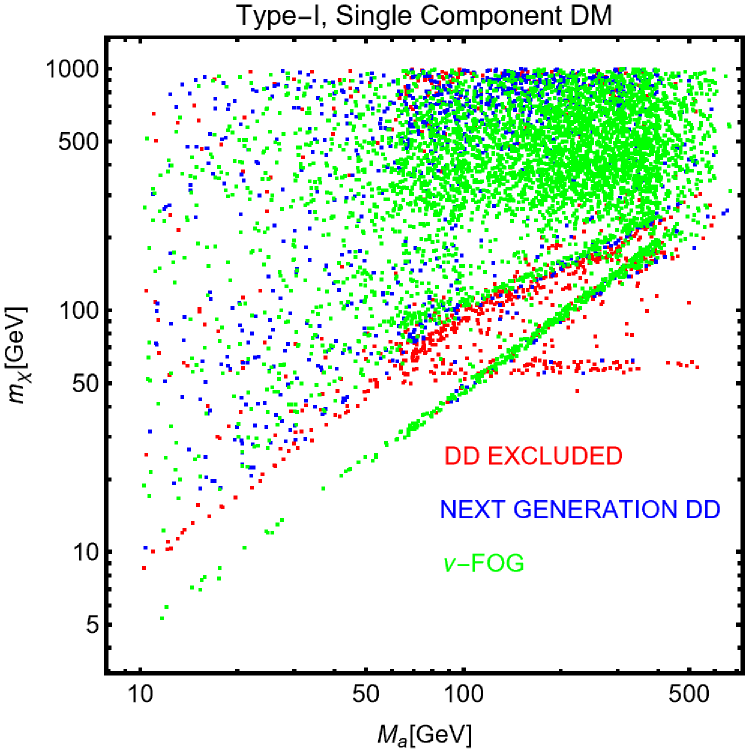}
    \includegraphics[width=0.49\linewidth]{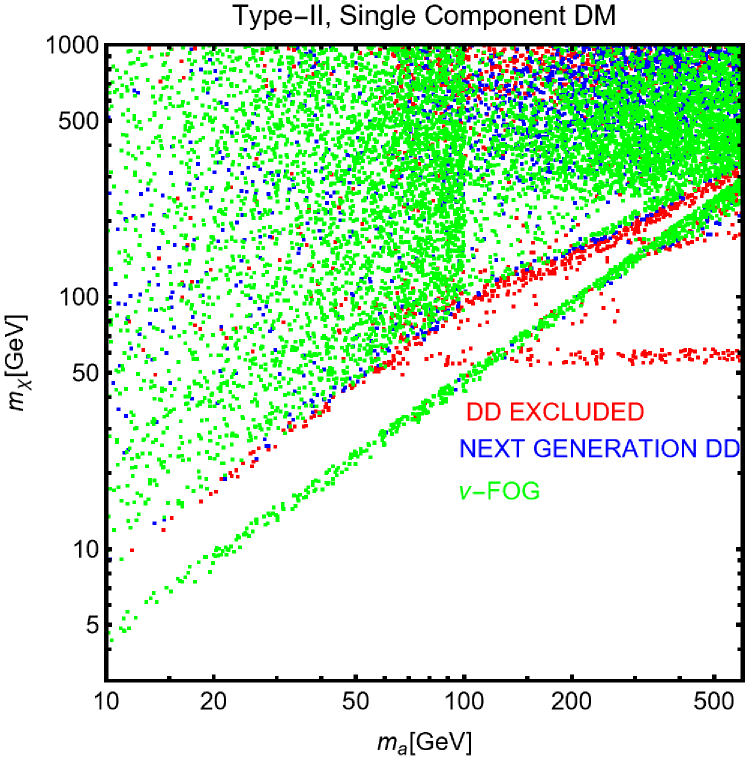}\\
    \includegraphics[width=0.49\linewidth]{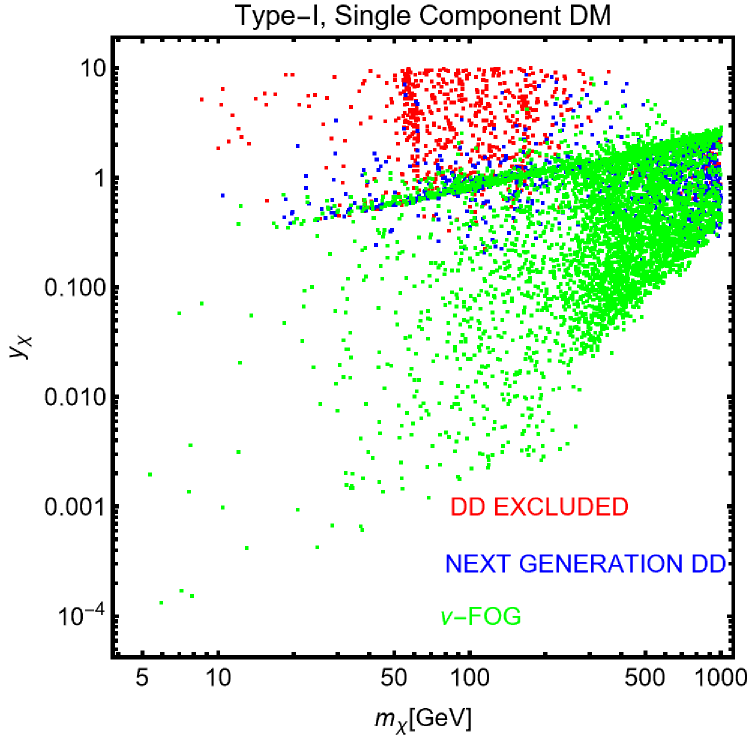}
    \includegraphics[width=0.49\linewidth]{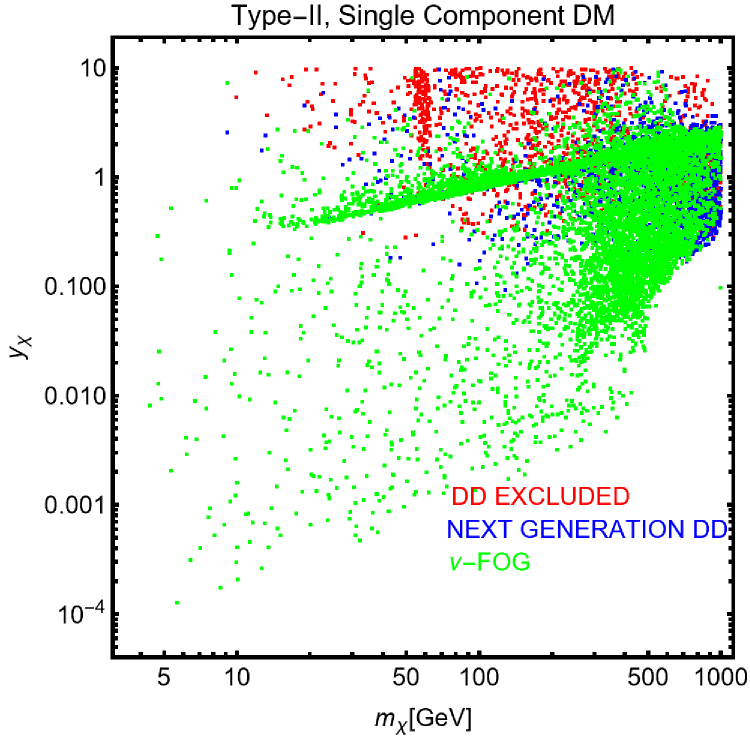}
    \caption{Model points of the 2HD+a complying, at the same time, with the requirement of the correct relic density, according to the conventional WIMP paradigm, and the present constraints from indirect detection. These model points are visualized, in the first row of the plot, in the $(M_a,m_\chi)$ plane while, in the second row, are visualized in the $(m_\chi,y_\chi)$ plane. In each panel the red points correspond to the parameter assignments excluded by current direct detection limits, the blue points to the assignments compatible with current limits but possibly reachable by next generation null results while, finally, the green point correspond to parameter assignments lying inside the neutrino fog.}
    \label{fig:plotthsum}
\end{figure}

Before discussing, in the next section, the case of two DM component, we provide below a summary of our findings up to now.
Figure~\ref{fig:plotthsum} summarises the impact of the combined DD and ID
constraints on the 2HD+a in the single-component regime ($\sin\theta \neq 0$).

\begin{figure}
    \centering
    \includegraphics[width=0.49\linewidth]{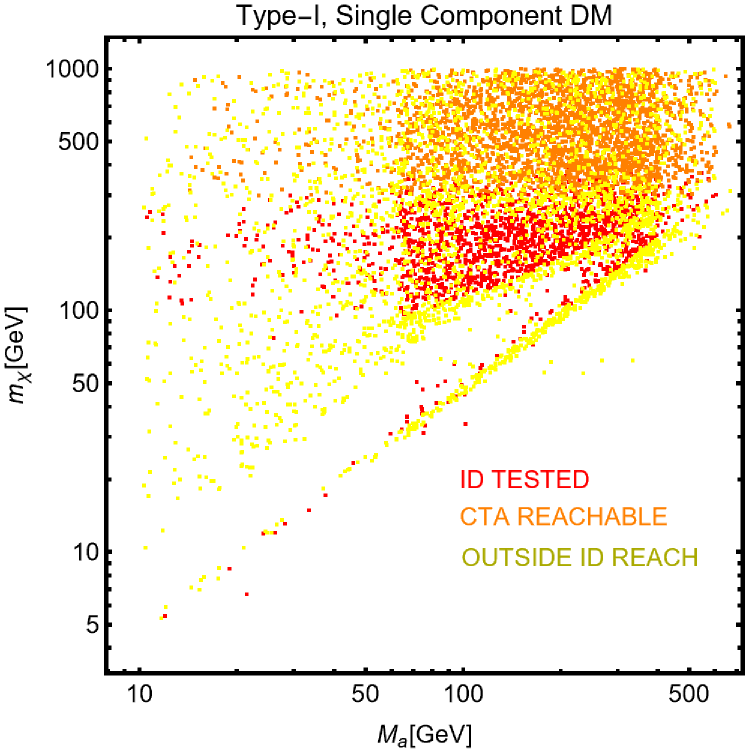}
     \includegraphics[width=0.49\linewidth]{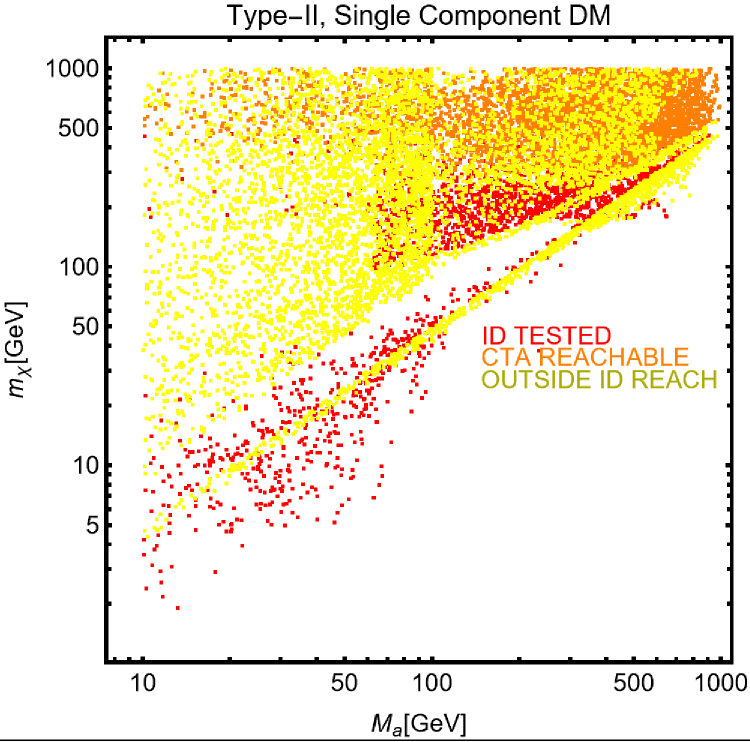}\\
     \includegraphics[width=0.49\linewidth]{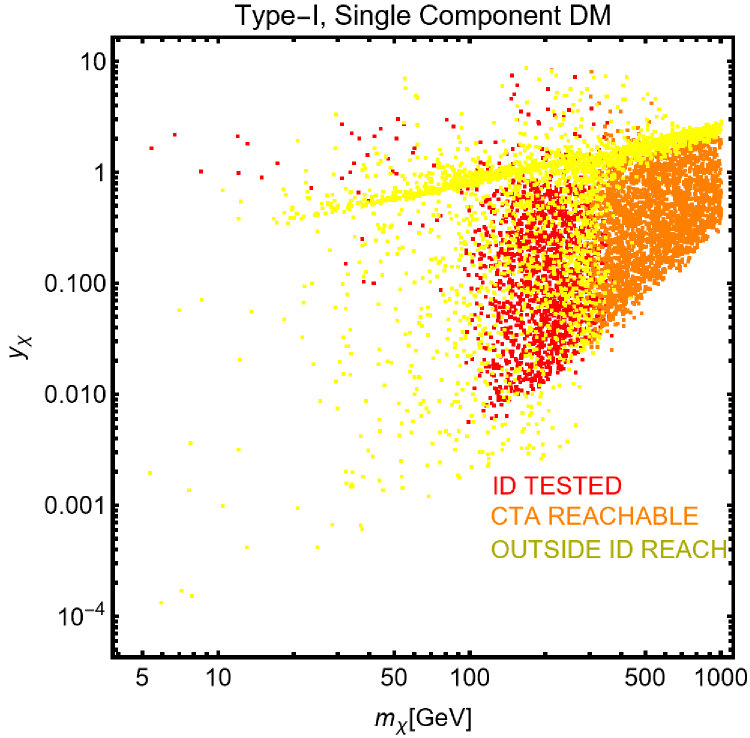}
     \includegraphics[width=0.49\linewidth]{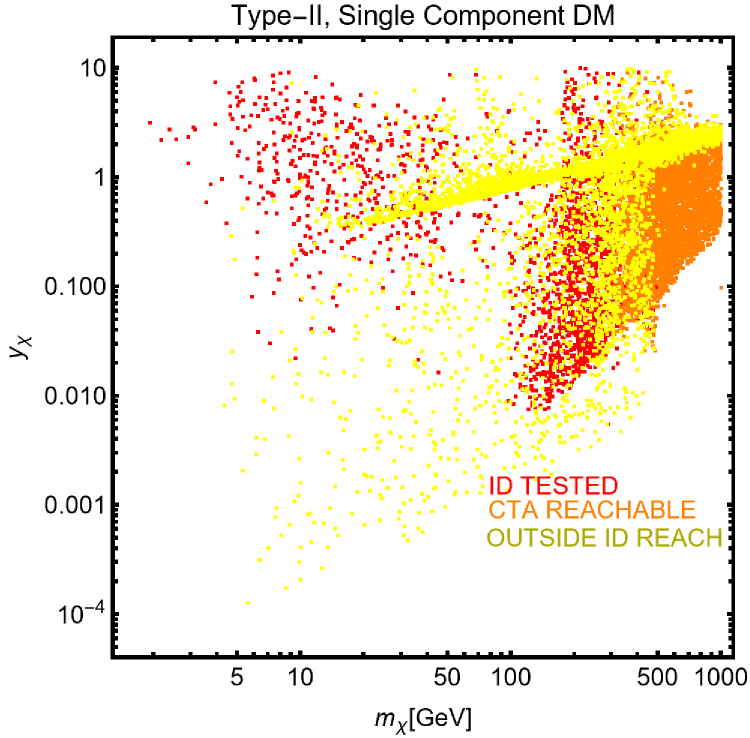}
    \caption{Model points of the 2HD+a complying, at the same time, with the requirement of the correct relic density, according to the conventional WIMP paradigm, and the present constraints from direct detection. These model points are visualized, in the first row of the plot, in the $(M_a,m_\chi)$ plane while, in the second row, are visualized in the $(m_\chi,y_\chi)$ plane. In each panel the red points correspond to the parameter assignments excluded by current $\gamma$-ray searches by FERMI-LAT, the orange points to the assignments compatible with current limits but possibly excluded by negative signals in the near future by CTA while, finally, the yellow point corresponds to parameter assignments unconstrained by ID.}
    \label{fig:plotthsumID}
\end{figure}

Several features stand out from Figs.~\ref{fig:plotthsum} and \ref{fig:plotthsumID}.
The bulk of viable model points lie when $m_\chi > M_a$. As already pointed out, the neutrino fog region (green points) is naturally populated for $m_\chi \gtrsim 100\,\mbox{GeV}$, when the annihilation into $ha$ final states is kinematically possible, while at low DM masses the neutrino fog is reached by fine-tuning the model parameters to suppress the $\lambda_{haa}$ coupling. In this last case, the relic density is achieved mostly via the $\bar \chi \chi \rightarrow a a$ annihilation process. In this setup the requirement of the correct DM relic translates into a very precise prediction for the DM Yukawa coupling $y_\chi$, evidenced by a narrow strip of green points in the $(m_\chi,y_\chi)$ plane. An additional viable region extends to low DM mass, where the relic density
can be satisfied only near the $s$-channel pole $m_\chi \simeq \frac12 M_a$,
represented by the narrow diagonal band in the $(M_a,m_\chi)$. 
Finally, an additional strip appears near
$m_\chi \simeq \frac12 M_h \simeq 63\,\text{GeV}$, arising from the loop-induced
$\bar\chi\chi h$ coupling, but this pole region is excluded by DD constraints
in both Type-I and II configurations. 
DM indirect detection represents an effective complementary probe, in particular in the high mass regions which will be accessible to the next future searches by CTA (orange points in Fig. \ref{fig:plotthsumID}). A bulk of points not accessible to both DD and ID is nevertheless present. This corresponds to the parameter assignments in which $\lambda_{haa}$ is fine-tuned to be very suppressed and at the same time the correct relic density is achieved via the p-wave suppressed annihilation process into $aa$ final states. Improved measures/constraints on the Higgs width of the represent a crucial tool to test such regions of the parameter space.

\subsection{\boldmath$\sin\theta = 0$ regime: two-component dark matter}
\label{sec:2hdma_zero}

In the extreme limit $\sin\theta = 0$ the mass eigenstate $a$ is a pure
${\rm SU(2)_L}$ singlet and does not couple directly to SM fermions via the
pseudoscalar portal.
If $M_a < 2m_\chi$, the state $a$ is cosmologically stable and constitutes
a second DM component alongside $\chi$. For what the relic density is concerned, this scenario is consistently treated by the micrOMEGAs package which, in this setup can solve a system of coupled Boltzmann's equation for the $\chi$ and $a$ DM components. In this case, the sum of the abundances of the two components will be confronted with the observed value of the DM relic density. For the range of parameters considered in the scan, despite the absence of a direct fermionic coupling, $a$ remains coupled to
the CP-even Higgs sector through $\lambda_{h,H aa}$, so it can annihilate
efficiently into SM fermions and gauge bosons via virtual $h/H$ exchange. For the fermionic component $\chi$, even with the portal coupling switched off,
thermal equilibrium is maintained through the processes $\chi\chi \to aa$ and
$\chi\chi \to a\,h(H)$. In this setup such process will act as well a conversion processes between the two DM components and are consistently treated in the system of Boltzmann's equations. In summary, the scenario under scrutiny can be treated as two component WIMP scenario \cite{Arcadi:2016kmk}.

\begin{figure}
    \centering
    \includegraphics[width=0.49\linewidth]{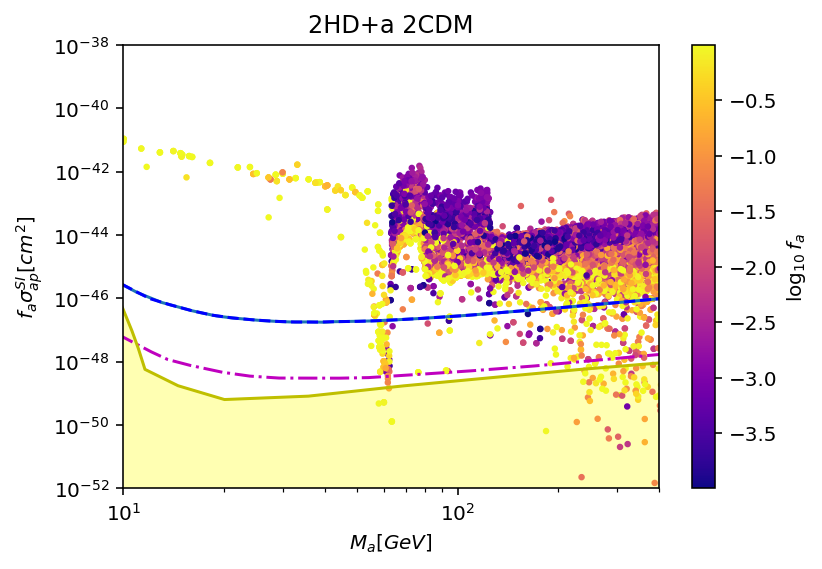}
     \includegraphics[width=0.49\linewidth]{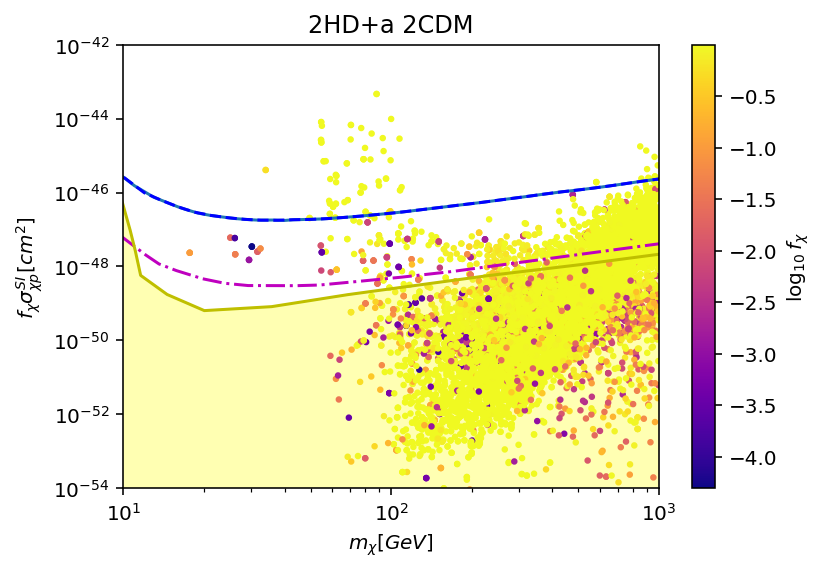}\\
    \includegraphics[width=0.49\linewidth]{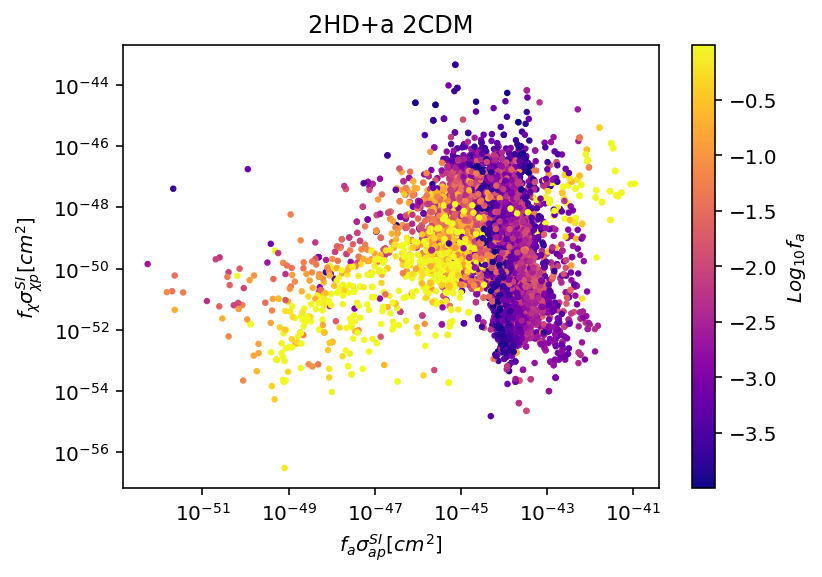}
    \caption{Parameter scan of the 2HD+a in the two-component DM (2CDM) freeze-out
    regime. \textit{Upper left}: model points compatible with the correct relic density
    in the $(m_\chi, f_a\sigma_{ap}^{\rm SI})$ plane; the model points above the blue line have component-weighted pseudoscalar cross-section residing above current constraints and, consequently, can be regarded as rule out. Color encodes the fraction $f_a$ of the total relic density
    carried by the pseudoscalar component.
    \textit{Upper right}: the same points in the $(m_\chi, y_\chi)$ plane for the
    fermionic component; $f_\chi$ approaches unity across most of the space,
    confirming that $\chi$ dominates the relic density while $a$ is a subdominant
    but stable companion. Color encodes $f_\chi$.
    \textit{Lower panel}: the same points in the $(f_a\sigma_{ap}^{\rm SI},\,
    f_\chi\sigma_{\chi p}^{\rm SI})$ plane, making the joint DD reach explicit;
    color encodes $f_a$. Model points in the left bottom corner of the panel have component weighted cross-sections below the neutrino-fog for both pseudoscalar and fermionic DM components. These points are   inaccessible
    to conventional direct detection across the thermally produced parameter space.}
    \label{fig:plot_2HDMa_2CDM_FO}
\end{figure}

Moving to direct detection, the most notable difference with respect to the $\sin\theta \neq 0$ case is the presence of scattering of the pseudoscalar component with nucleons, occurring at tree level via $t$-channel $h/H$ exchange:
\begin{equation}
\label{eq:sigma_ap}
\sigma_{ap}^{\rm SI} = \frac{\mu_{ap}^2}{\pi}
\frac{m_p^2}{M_a^2 v_h^2}
\left|\sum_q f_q^p
\left(\frac{\lambda_{haa}}{M_h^2} + \frac{\lambda_{Haa}}{M_H^2}\right)
\right|^2,
\end{equation}
where $\mu_{ap}$ is the $a$--proton reduced mass and the sum runs over
light quarks weighted by their hadronic matrix elements $f_q^p$.
Additionally, the effective loop-induced $\bar\chi\chi h$ coupling that generates
the SI cross-section in the single-component case persists at $\sin\theta\!=\! 0$
and provides a residual contact with nucleons.
The treatment of DD of 2CDM requires, in general, a dedicated statistical treatment, see e.g. \cite{Profumo:2009tb}. Here we will limit to individually compare, the component weighted cross-section $f_i \sigma_{ip}^{\rm SI}$ with the experimental limit. As evident from figure \ref{fig:plot_2HDMa_2CDM_FO} such weighted cross-sections differ by several orders of magnitude for most of the model points so that DD phenomenology is mostly determined by the pseudoscalar component. 
The detection prospects in the 2CDM regime are displayed in
Fig.~\ref{fig:plot_2HDMa_2CDM_FO}.
Since the Yukawa configurations have negligible impact on the 2CDM
phenomenology (there is no tree-level $a$--fermion coupling), we show results
without separating Type-I and II.

\begin{figure}
    \centering
    \includegraphics[width=0.43\linewidth]{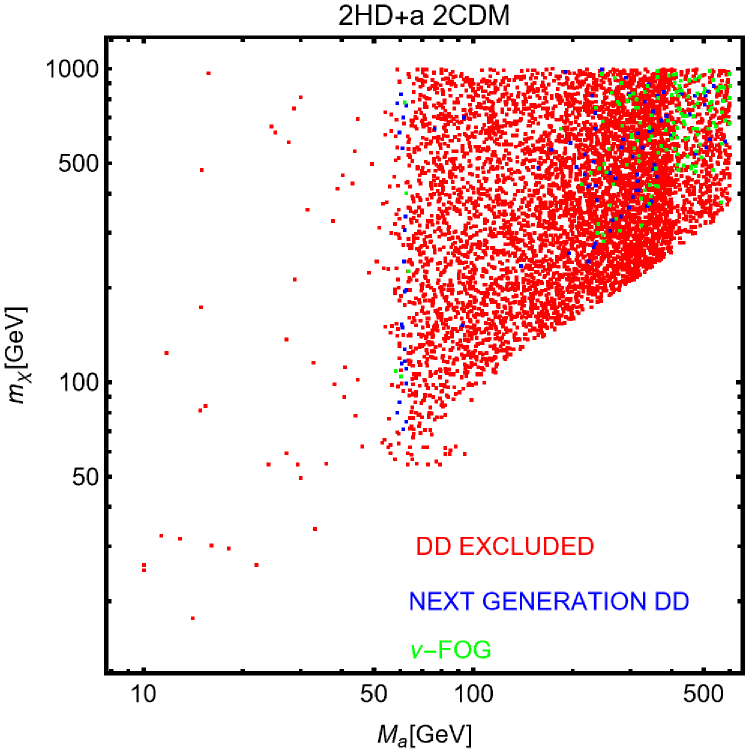}
    \includegraphics[width=0.43\linewidth]{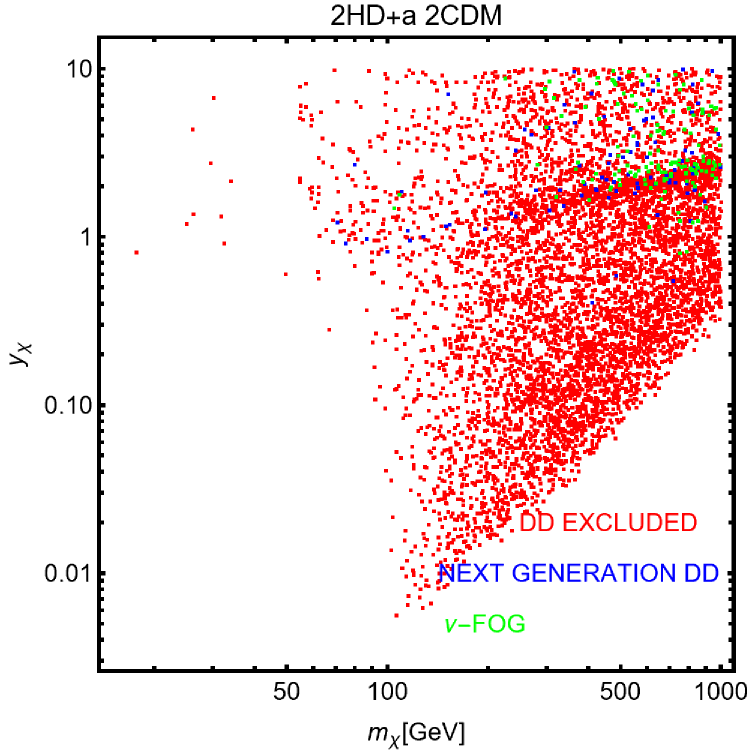}
    \caption{The same as Fig. \ref{fig:plotthsum} but in the 2CDM regime.}
    \label{fig:pp2CDM}
\vspace*{-3mm}
\end{figure}

The upper-left panel of Fig.~\ref{fig:plot_2HDMa_2CDM_FO} shows that the
component-weighted cross-section $f_a\,\sigma_{ap}^{\rm SI}$ resides above the experimental limits. This is consequence of tree-level nature of the SI interactions. One notice in particular that the model points with low $f_a$ correspond to the highest values of the scattering cross-section since higher couplings are needed to suppress the relic density of the pseudoscalar component. The upper-right panels shows that the weighted cross-section of the fermionic component mostly resides below the experimental limits with the majority of points being inside the neutrino fog. The bottom panel of the figure shows that it is possible anyway to have model configurations for which both weighted cross-section are suppressed below detection limits. These correspond to the case in which the coupling $\lambda_{haa}$ is fine-tuned to be suppressed. If $M_a \gtrsim 100\,\mbox{GeV}$ (see upper panel of Fig. \ref{fig:plot_2HDMa_2CDM_FO}) the correct relic density nevertheless achieved via interactions involving the heavy CP-even state $H$. This does not occur at lighter masses of $a$ where one would need very high value of the coupling $y_\chi$, again implying a too high scattering cross-section for the pseudoscalar component.

We conclude our analysis by analogous study of the parameter space is performed in Fig. \ref{fig:pp2CDM}. As evident, the vast majority of the parameter space is already excluded by current direct detection constraints as the pseudoscalar DM features tree-level induced interactions with nucleons. New avenues for DM production, as for example the non-thermal production of at least one of the DM components, see e.g. \cite{Belyaev:2022qnf,Amiri:2025ras}, should be considered.

%%%%%%%%%%%%%%%%%%%%%%%%%%%%%%%%%%%%%%%%%%%%%%%%%%
\subsection{LHC signatures}
\label{sec:2hdma_lhc}

The 2HD+a predicts a broad range of LHC signatures, primarily through the
resonant production of the heavier $H$ and $A$  bosons which subsequently
decay into lighter states.
In the regime $\sin\theta \gtrsim 10^{-3}$ the light pseudoscalar $a$ decays
promptly at the interaction point into SM fermion or gauge boson pairs.

\begin{figure}
    \centering
    \includegraphics[width=0.32\linewidth]{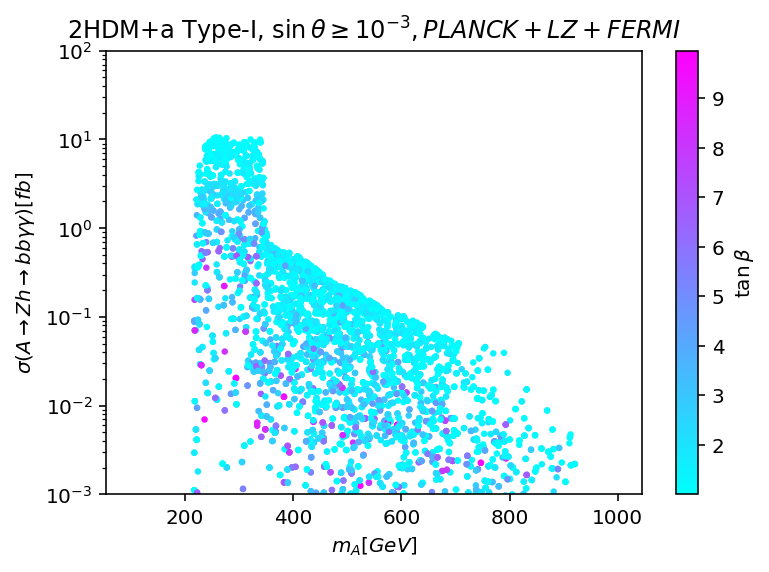}
    \includegraphics[width=0.32\linewidth]{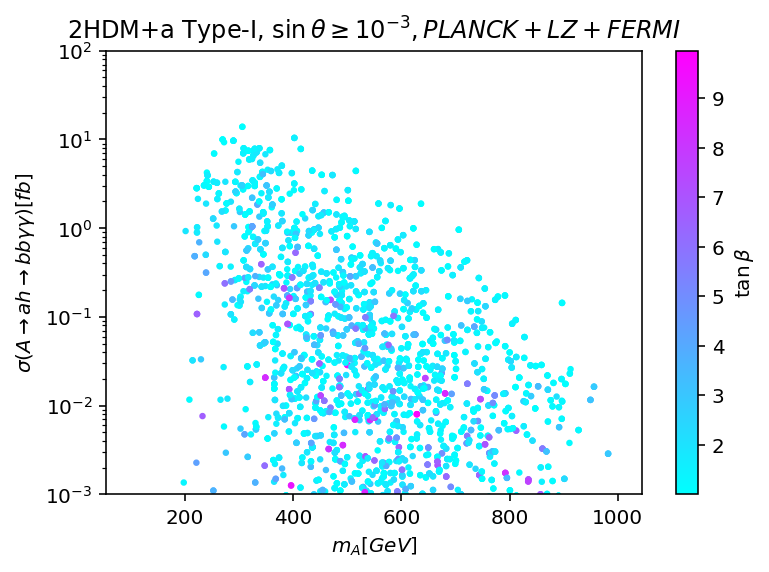}
    \includegraphics[width=0.32\linewidth]{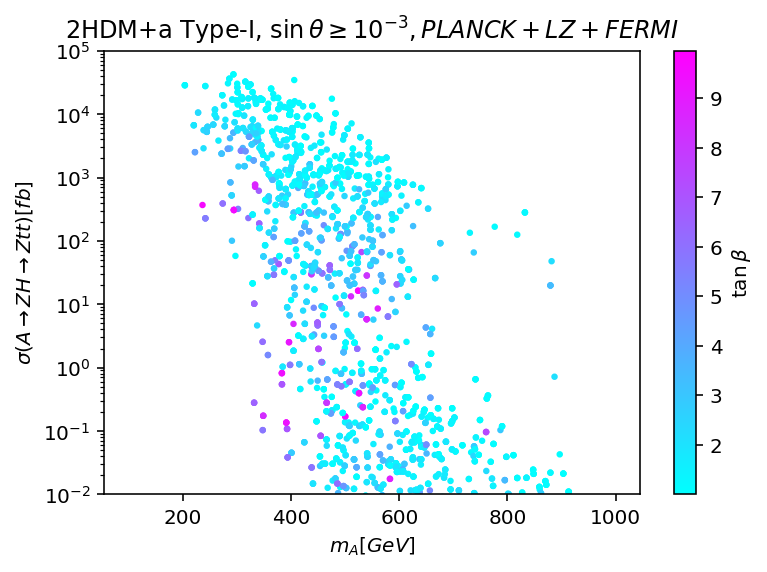}\\
    \includegraphics[width=0.32\linewidth]{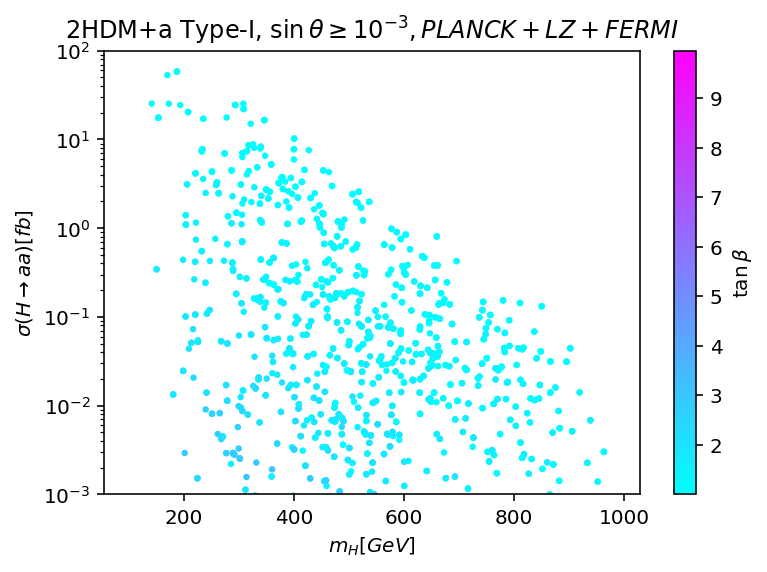}
    \includegraphics[width=0.32\linewidth]{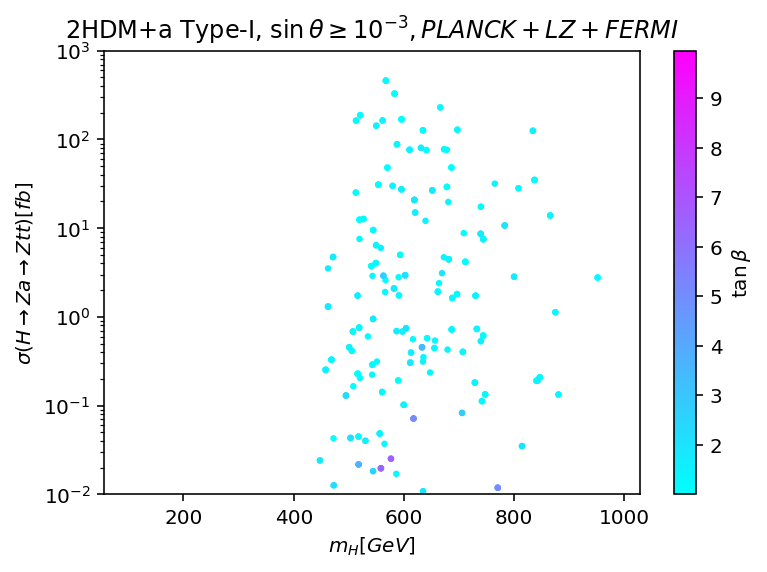}
     \includegraphics[width=0.32\linewidth]{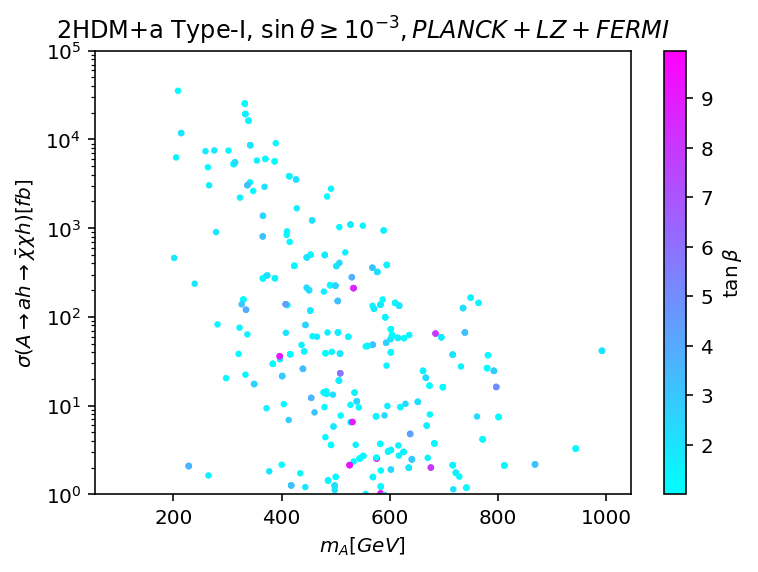}
    \caption{Production cross-section for a series of resonance production processes, which are customarily used to probe the 2HD+a at colliders. In each panel, the points correspond to parameter assignment complying with all the constraints, including DM direct and indirect detection, considered in this work. The color pattern of the points correspond to $\tan\beta$.}
    \label{fig:pTypILHC}
\end{figure}

\begin{figure}
    \centering
    \includegraphics[width=0.33\linewidth]{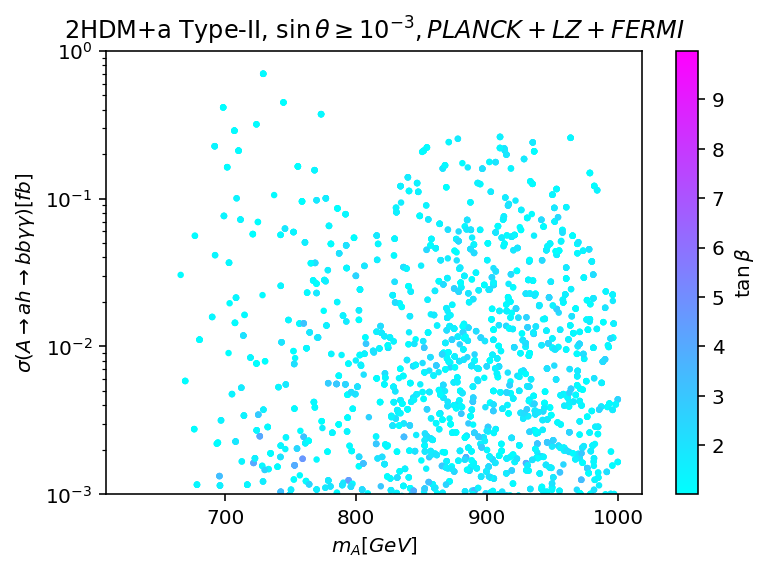}
    \includegraphics[width=0.33\linewidth]{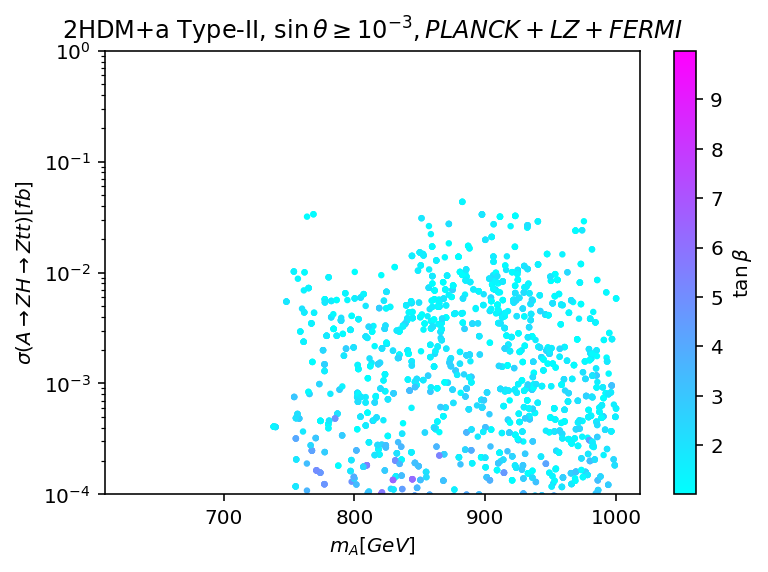}\\
    \includegraphics[width=0.32\linewidth]{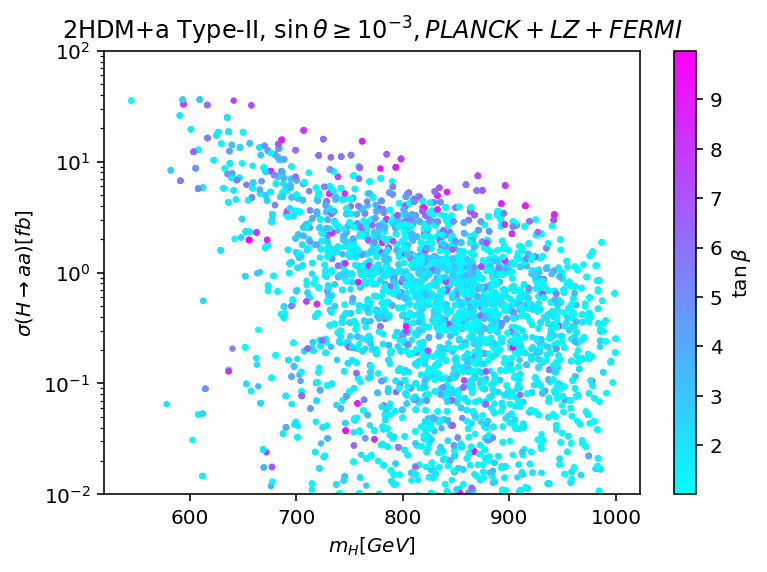}
    \includegraphics[width=0.32\linewidth]{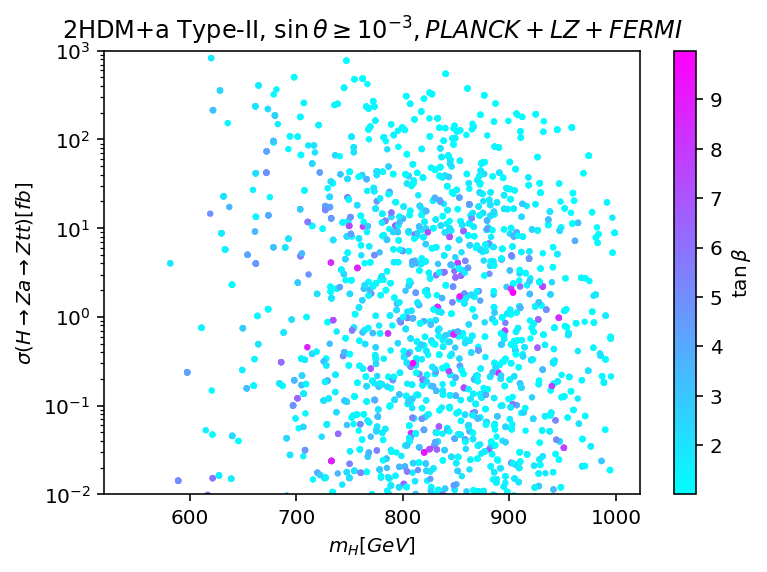}
    \includegraphics[width=0.32\linewidth]{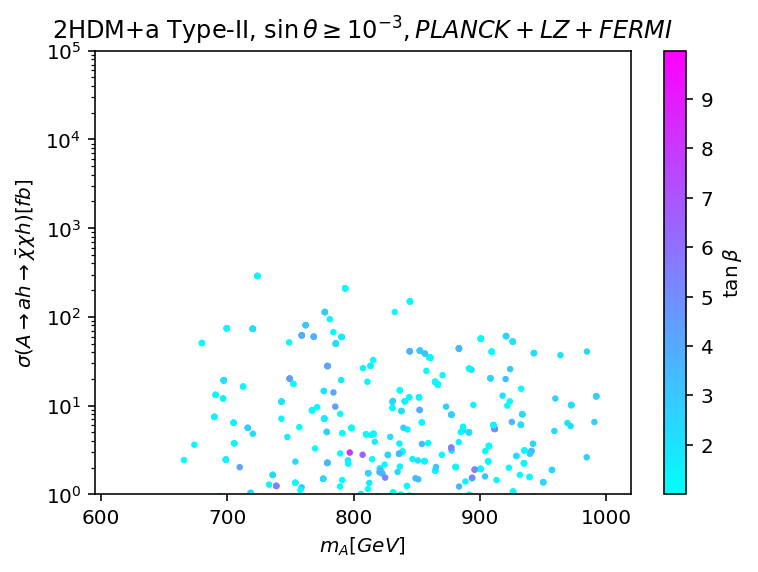}
    \caption{The same as Fig. \ref{fig:pTypILHC} but for the Type-II model. The panel relative to the $pp \rightarrow A \rightarrow Zh$ process is not present as the coupling $AZh$ is zero in the alignment limit.}
    \label{fig:pTypIILHC}
\end{figure}

Figures~\ref{fig:pTypILHC} and \ref{fig:pTypIILHC} display the production
cross-sections for a representative set of signatures, for Type-I and Type-II
respectively. All the cross-section have been computed by rescaling by suitable $\cos^2 \theta$ or $\sin^2 \theta$ factors the cross-section determined by the package SusHi \cite{Harlander:2012pb,Harlander:2016hcx} for the FCNC preserving 2HDMs.
In each panel, only model points satisfying all constraints considered in
this work are shown (theoretical, electroweak precision, flavor, relic density,
LZ, FERMI-LAT), and the color encodes $\tan\beta$.
The signatures considered include processes already searched for at the LHC:
$pp \to A \to ah \to \bar{b}b\gamma\gamma$ \cite{CMS:2023boe},
$pp \to H \to aa$ \cite{ATLAS:2026epu,ATLAS:2024nnm}, $pp \to A \to h a(\to \bar \chi \chi)$, and
$pp \to H \to aZ \to Zt\bar{t}$ \cite{ATLAS:2023zkt}.
We also include complementary channels not directly involving $a$:
$pp \to A \to Zh \to \bar{b}b\gamma\gamma$ and
$pp \to A \to ZH \to Zt\bar{t}$.

Comparing the two Yukawa types, the Type-II scenario covers a narrower mass
range: the $b \to s\gamma$ constraint forces $M_{H^\pm} \gtrsim 800\,\text{GeV}$,
which, combined with perturbative unitarity and electroweak precision bounds,
pushes the masses of all heavier CP-even and charged states to high values.
In both scenarios sizable production cross-sections are concentrated at
$\tan\beta \leq 3$.
The $AZh$ final state is absent in Type-II as the coupling $AZh$ is zero in the
alignment limit.

As evident from the plots, the values of the production cross-section span several orders of magnitude. A systematic assessment of the detection prospects of the scenario under scrutiny is beyond the scopes of present work, mostly focussed on DM phenomenology. Furthermore a comparison with current LHC searches is often not trivial as experimental results are often interpreted in terms of specific benchmarks. 

We can nevertheless provide a concrete sense of experimental reach, noticing, for example, that in the Type-I
scenario the $pp \to A \to ah \to \bar{b}b\gamma\gamma$ cross-section
reaches $\mathcal{O}(10\,\text{fb})$ for $M_A \sim 400$--$700\,\text{GeV}$
at $\tan\beta \lesssim 3$, already in the reach of LHC searches \cite{CMS:2023boe}.
A portion of the viable DM parameter space is then already within the current LHC reach and an even larger portion can be covered by HL-LHC, 
On similar footing, our scan evidenced a portion of points with cross-sections of the $pp \to A \to ZH \to Zt\bar{t}$ processes of the order or above current sensitivity, see e.g. \cite{ATLAS:2023zkt}.

In all cases  the viable points shown in Figs.~\ref{fig:pTypILHC} and
\ref{fig:pTypIILHC} satisfy the full set of DM and precision constraints, so
the cross-sections displayed represent genuinely unconstrained predictions
for future searches rather than post-hoc benchmarks.

\begin{figure}
    \centering
    \includegraphics[width=0.45\linewidth]{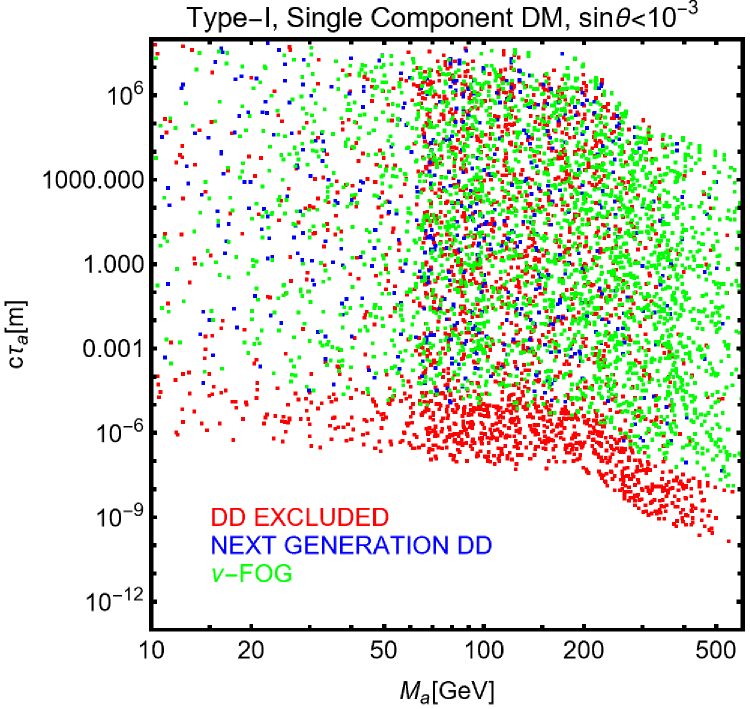}
    \includegraphics[width=0.41\linewidth]{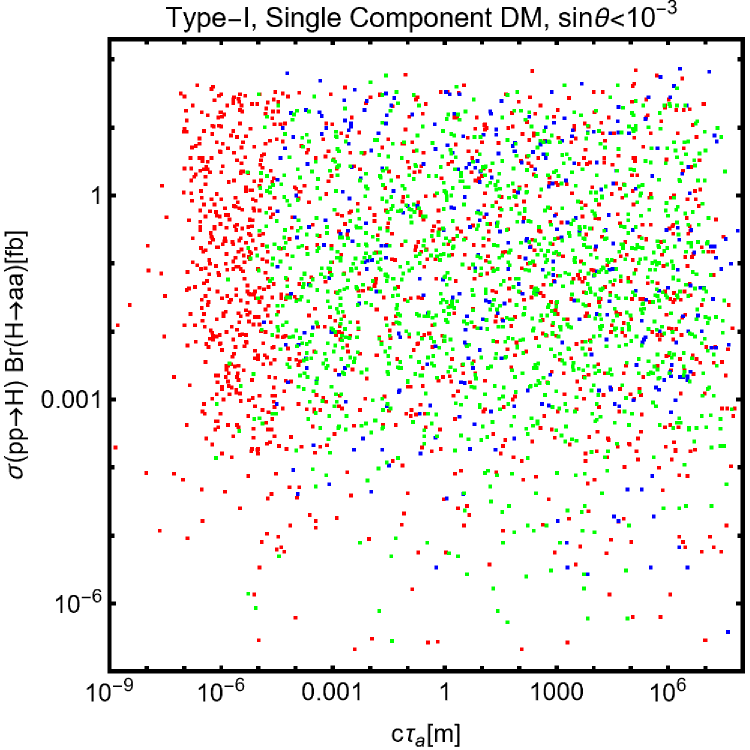}\\
    \includegraphics[width=0.45\linewidth]{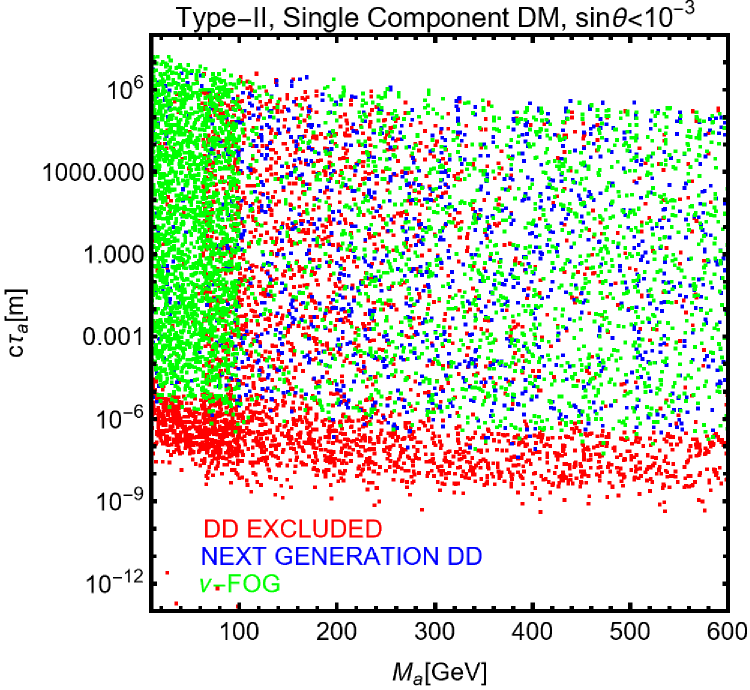}
    \includegraphics[width=0.41\linewidth]{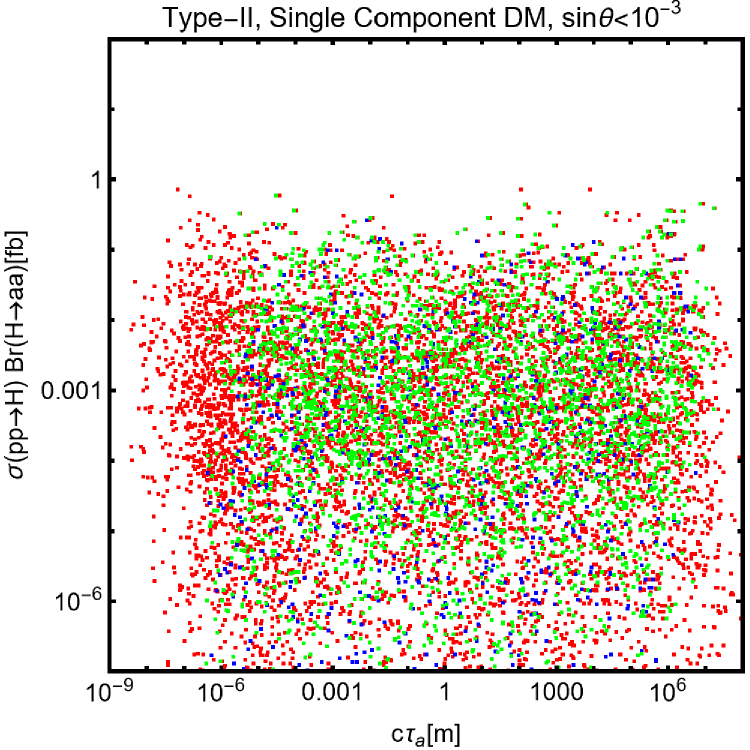}
    \vspace*{-2mm}
    \caption{{\it Left column}: Points of the parameter scan, in the $\sin\theta < 10^{-3}$ regime, in the $(M_a,c\tau_a)$ plane. All the points displayed comply with the correct DM relic density and the present bounds from indirect detection. Points marked in blue are also compatible with present constraints from direct detection while the ones marked in green have scattering cross-section lying inside the neutrino fog. {\it Right column}: Value of the cross-section of the $pp \rightarrow H \rightarrow aa$ process as function of $c \tau_a$. The points, as well as the color code, are the same as the ones displayed in the left column.}
    \label{fig:pLLPLHC}
\vspace*{-5mm}
\end{figure}

In the long-lived regime $\sin\theta \! < \! 10^{-3}$, with $M_a \!< \!2m_\chi$ as
preferred by DM constraints, the state $a$ is long-lived at LHC scales.
The primary collider signature is then the production of pairs of displaced
hadronic jets from $a$ decays, following the resonant production $pp \! \to \! h,H \! \to \! aa$.
Figure~\ref{fig:pLLPLHC} shows the parameter-scan points in the $(M_a,\, c\tau_a)$
plane (left column) and the corresponding $pp \to H \to aa$ cross-section
as a function of $c\tau_a$ (right column), for Type-I (upper row) and Type-II
(lower row).
Points inside the neutrino fog (green) have $c\tau_a$ spanning a wide range,
many within the acceptance of existing ATLAS displaced-jet searches
\cite{ATLAS:2018tup,ATLAS:2019qrr,ATLAS:2019jcm,ATLAS:2022gbw,ATLAS:2022zhj},
reinforcing the complementarity between DM and collider probes.

\section{Conclusions}
\label{sec:conclusions}

The neutrino fog marks the frontier where conventional WIMP direct detection
loses its sharpest edge.
As tonne-scale liquid xenon experiments approach this barrier, a central question
becomes whether thermally produced WIMPs can naturally populate the sub-floor
cross-section region without sacrificing viable relic density or detectable indirect
signatures.
We have shown, through a systematic parameter scan of the 2HD+a model,
that the answer is yes:  the model predicts
a structured, testable phenomenology precisely in this regime.

In the 2HD+a model the phenomenology is rich and hierarchical.
In the standard mixing regime $\sin\theta \geq 10^{-3}$
(section~\ref{sec:2hdma_high}), the loop-induced SI cross-section already
allows a significant fraction of relic-density-compatible points to lie inside the
neutrino fog, particularly at $m_\chi \gtrsim 100\,\text{GeV}$ where the
$ha$ annihilation channel governs freeze-out.
At strongly suppressed mixing $\sin\theta < 10^{-3}$
(section~\ref{sec:2hdma_low}), the $\bar{f}f$ annihilation channel closes and
freeze-out proceeds almost exclusively via $ha$ and $Ha$; the SI cross-section
is entirely loop-induced and for the smallest values of $|\sin\theta|$ falls
comfortably inside the fog.
The same suppression of $\sin\theta$ lengthens the $a$ lifetime to macroscopic
scales, producing a population of displaced-jet signatures at the LHC that
is correlated with the neutrino-fog parameter space: a substantial fraction of
the green (sub-floor) points in Fig.~\ref{fig:2HDMa_tau} have $c\tau_a$ values
within the acceptance of existing ATLAS displaced-jet searches
\cite{ATLAS:2018tup,ATLAS:2019qrr,ATLAS:2019jcm,ATLAS:2022gbw,ATLAS:2022zhj},
and production cross-sections in the $10^{-3}$--$1\,\text{fb}$ range accessible
to LHC Run-3 and the HL-LHC.
In the extreme limit $\sin\theta = 0$ (section~\ref{sec:2hdma_zero}),
a two-component WIMP scenario emerges: the pseudoscalar $a$ becomes stable and
scatters on nucleons via tree-level $h/H$ exchange. Among the three, this appears to be the most constrained scenario because of the tree-level induced coupling between the pseudoscalar DM component and nucleons.

Several experimental implications follow directly from these results.

\textit{Direct detection.}
The WIMP parameter space does not end at the neutrino fog:
a substantial and theoretically motivated region of the 2HD+a lies below
current LZ sensitivity and within the neutrino fog.
This region is not unreachable in principle; it requires statistical techniques to
discriminate DM signals from the CE$\nu$NS background \cite{OHare:2021utq},
and tonne-scale detectors with directional sensitivity or improved neutrino flux
measurements can in principle recover sensitivity there.
The full exploitation of XLZD \cite{XLZD:2024nsu} and its successors is therefore
directly motivated by the models studied here.

\textit{Indirect detection.}
The DD/ID complementarity is a robust prediction of the model.
The summary plots (figs.~\ref{fig:plotthsum} and \ref{fig:plotthsumID})
show that the green (neutrino-fog) points are predominantly unconstrained by current
FERMI-LAT searches but within projected CTA reach for $m_\chi \lesssim 500\,\text{GeV}$.
This means that CTA, rather than the next generation of xenon experiments,
may be the instrument that first probes the sub-floor WIMP parameter space
established here, particularly for DM masses in the $100$--$500\,\text{GeV}$ range.

\textit{LHC.}
The 2HD+a LHC signatures surveyed in section~\ref{sec:2hdma_lhc} show
complementarity of a different kind.
In the prompt regime, sizable production cross-sections for $pp \to A \to ah$,
$pp \to H \to aa$, and related channels are achievable at $\tan\beta \leq 3$,
even for parameter assignments inside the neutrino fog.
In the long-lived regime, the same loop suppression that hides DM from direct
detection also lengthens $\tau_a$, generating displaced hadronic signatures whose
production rate is correlated with the sub-floor DD cross-section.
This correlation makes the combination of xenon detector non-observations and
LHC displaced-jet searches a particularly powerful probe of the low-$\sin\theta$
parameter space.

\textit{Invisible Higgs width and $h \to aa$.}
The constraint $\text{Br}(h \to \text{inv.}) \leq 0.11$ eliminates the kinematic window of $h\rightarrow aa$ decay unless the $\lambda_{haa}$ is suppressed via a fine-tuning of the model parameters.
Conversely, this means that any future signal in $h \to aa$ searches at the
HL-LHC \cite{ATLAS:2025qyn,ATLAS:2024nnm,CMS:2025hjt} would constitute
evidence against the standard 2HD+a freeze-out scenario studied here,
providing a sharp discriminant.

Two broader lessons emerge from this work.
First, pseudoscalar mediators are not merely a mechanism for evading direct
detection constraints: they predict a specific pattern of complementarity
between direct detection, indirect detection, and collider searches that is
testable with experiments already under construction or approved.
Second, the neutrino fog is best understood not as a barrier but as a
threshold beyond which the multi-channel nature of DM phenomenology becomes
essential.
Neither direct detection alone nor indirect detection alone nor LHC alone
can fully characterize the parameter space studied here; the combination of
all three is necessary and, taken together, largely sufficient to probe the
model across the mass range $10$--$1000\,\text{GeV}$.

A natural extension of this work is the inclusion of CP-violating interactions
directly in the 2HD+a scalar sector, which would introduce tree-level
mixing between the CP-even and CP-odd mass eigenstates and potentially generate
EDM signals alongside the DD/ID complementarity studied here.
Another avenue is to relax the thermal freeze-out assumption: in the limit
$\sin\theta \to 0$ the DM relic abundance could in principle be set by
non-thermal mechanisms, opening additional parameter space inside the
neutrino fog that deserves dedicated investigation.

%%%%%%%%%%%%%%%%%%%%%%%%%%%%%%%%%%%%%%%%%%%%%%%%%%
\subsection*{Acknowledgments}
AD was supported by the Spanish grant PID2021-128396NB-I00. 
SP was supported by the U.S.\ Department of Energy, Office of Science, Office of High Energy Physics, under Award Number DE-SC0010107.

%%%%%%%%%%%%%%%%%%%%%%%%%%%%%%%%%%%%%%%%%%%%%%%%%%

%\newpage

\bibliographystyle{utphys} % for Physics and Mathematics articles
\bibliography{bibfile}
\end{document}